\documentclass[10pt,a4paper,reqno]{amsart}
\usepackage{acronym}
\usepackage{amsfonts}
\usepackage{amsmath}
\usepackage{amssymb}
\usepackage{amsthm}
\usepackage{bm}
\usepackage{braket}
\usepackage{cite}
\usepackage{color}
\usepackage{geometry}
\usepackage{graphicx}
\usepackage{hyperref}
\usepackage{mathrsfs}
\usepackage{mathtools}
\usepackage{setspace}
\usepackage{stmaryrd}
\usepackage{subcaption}
\usepackage{tikz}

\usepackage{epsfig}
\usepackage{setspace}
\usepackage{booktabs}
\usepackage{threeparttable}
\usepackage{diagbox}
\usepackage{epstopdf}

\newgeometry{top=2cm,bottom=2cm,outer=1.5cm,inner=1.5cm}
\newcommand{\bse}{\begin{subequations}}
\newcommand{\ese}{\end{subequations}}

\numberwithin{equation}{section}
\DeclareSymbolFont{largesymbols}{OMX}{yhex}{m}{n}
\DeclareMathAccent{\Widehat}{\mathord}{largesymbols}{"62}

\title[A two-stage PINN method based on conserved quantities and applications in localized wave solutions]{A two-stage physics-informed neural network method based on conserved quantities and applications in localized wave solutions}
\author{Shuning Lin}
\address[SL]{School of Mathematical Sciences, Shanghai Key Laboratory of Pure Mathematics and Mathematical Practice, and Shanghai Key Laboratory of Trustworthy Computing \\
East China Normal University \\ Shanghai 200241 \\ China}
\author{Yong Chen$^*$}
\address[YC]{School of Mathematical Sciences, Shanghai Key Laboratory of Pure Mathematics and Mathematical Practice, and Shanghai Key Laboratory of Trustworthy Computing \\
East China Normal University \\ Shanghai 200241 \\ China}
\address[YC]{College of Mathematics and Systems Science \\ Shandong University of Science and Technology \\ Qingdao 266590 \\ China}
\email{ychen@sei.ecnu.edu.cn}

\begin{document}

\begin{abstract}
With the advantages of fast calculating speed and high precision, the physics-informed neural network method opens up a new approach for numerically solving nonlinear partial differential equations. Based on conserved quantities, we devise a two-stage PINN method  which is tailored to the nature of equations by introducing features of physical systems into neural networks. Its remarkable advantage lies in that it can impose physical constraints from a global perspective. In stage one, the original PINN is applied. In stage two, we additionally introduce the measurement of conserved quantities into mean squared error loss to train neural networks. This two-stage PINN method is utilized to simulate abundant localized wave solutions of integrable equations. We mainly study the Sawada-Kotera equation as well as the coupled equations: the classical Boussinesq-Burgers equations and acquire the data-driven soliton molecule, M-shape double-peak soliton, plateau soliton, interaction solution, etc. Numerical results illustrate that abundant dynamic behaviors of these solutions can be well reproduced and the two-stage PINN method can remarkably improve prediction accuracy and enhance the ability of generalization compared to the original PINN method.


\end{abstract}
\maketitle

\section{Introduction}

Relying on the advantages of fast calculating speed and high precision, deep neural networks have developed rapidly and been applied widely in various fields, such as image recognition, speech recognition, natural language processing and so on. The neural network method also plays an important role in the area of scientific computing, especially in solving forward and inverse problems of nonlinear partial differential equations. As a major landmark, Raissi et al. proposed the physics-informed neural network (PINN) method \cite{Raissi2019}, which is one of the most powerful and revolutionary data-driven approaches. It aims to train neural networks to solve supervised learning tasks while respecting laws of physics described by nonlinear partial differential equations. On this basis, abundant significant physics-informed neural network frameworks, e.g. NSFnets \cite{NSFnets2021}, VPINNs \cite{VPINNs2019}, fPINNs \cite{fPINNs2019}, B-PINNs \cite{B-PINNs2021} and hp-VPINNs \cite{hp-VPINNs2021}, were devised and targeted at different application situations. This PINN methodology and its variants also have demonstrated extraordinary performance in approximating the unknown solutions \cite{Raissi2018,Yang2019,Mao2020}, data-driven discovery of partial differential equations \cite{RaissiKarniadakis2018,JagtagKawaguchi2020}, the research of extracting physical information from flow visualizations \cite{RaissiYazdani2020} and beyond \cite{Shin2020,ZhengQ2020}. In order to improve the performance of physics-informed neural networks, Jagtag et al. also proposed different ways of locally adaptive activation functions with slope recovery term and these methods are capable of accelerating the training process \cite{Jagtap2020}. Also noteworthy, many scholars have obtained a number of research results \cite{WangYan2021,Li2020, LiChen2020,Pu2021,PuLi2021,Peng2021}. Our group mainly focused on integrable equations possessing remarkable properties, such as the KdV equation, mKdV equation, nonlinear Schr\"odinger equation, derivative nonlinear Schr\"odinger equation (DNLS) and Chen-Lee-Liu equation \cite{Li2020, LiChen2020,Pu2021,PuLi2021,Peng2021}. By means of the PINN method, we reproduced abundant dynamic behaviors of data-driven solutions with regard to mentioned equations, including the breathing solution \cite{Pu2021}, rogue wave solutions \cite{Pu2021,PuLi2021}, rogue periodic wave \cite{Peng2021} and so on.

Currently, we are devoting to the research of integrable-deep learning algorithms, which aim to study integrable systems via the deep learning algorithm and further improve the neural network method with the advantages of integrable systems. First of all, numerous exact solutions can be obtained since integrable systems have outstanding properties. Therefore, it provides abundant samples for the PINN algorithm in reproducing dynamic behaviors of solutions. Secondly, due to the good properties of integrable systems such as abundant symmetry, infinite conservation laws and the Lax pair, as well as the mature methods for studying integrable systems including the Darboux transformation \cite{Wang2015, WangLi2015, Xu2017, XuChen2017, Xu2016}, the B\"acklund transformation \cite{Wahlquist1973, Hirota1977, Nakamura1979, Lu2012, Chen2020}, the Hirota bilinear method \cite{Zhang2017, Zhang2018} and the inverse scattering transformation \cite{Zhang2019, Ablowitz2016, Yu2019, Zhao2021}, we can combine these properties and methods with the PINN method to obtain more accurate numerical solutions. Finally, considering that integrable systems can describe physical phenomena such as the localized wave and turbulence \cite{Kupershmidt1985, Agafontsev2016, Agafontsev2020}, we can observe more physical phenomena with the aid of PINN method, which can not be obtained by classical methods.

 The two-stage physics-informed neural network method based on conserved quantities is proposed here. The specific way is that the original PINN is applied in stage one while in stage two we additionally introduce the measurement of conserved quantities into mean squared error loss to train neural networks. There are three motivations for this improved method. Above all, we intend to further improve integrable-deep learning algorithm and thus more features of physical systems are introduced into neural networks, such as conserved quantities considered in this paper. In the next place, our goal is to devise more targeted PINN method in solving nonlinear systems, especially integrable systems, which is tailored to the nature of equations by digging out more underlying information of the given equations. Last but not least, we aim to impose constraints from a global perspective considering that the loss functions in the original PINN method reflect the local constrains at certain points solely.

In this paper, we mainly consider nonlinear integrable equations: the Boussinesq-Burgers equations \cite{Kaup1975, Kupershmidt1985}, the classical Boussinesq-Burgers equations \cite{Date1978,Geng1999, Ito1984, Kawamoto1984, Gu1990} as well as the Sawada-Kotera equation \cite{Sawada1974, Caudrey1976}. Here, our improved PINN method is utilized to reproduce the dynamic behaviors of localized wave solutions for the above equations, such as the interaction solution, soliton molecule, M-shape double-peak soliton, etc.

 This paper is organized as follows. In Section. 2, we review the physics-informed neural network method for completeness and put forward the two-stage PINN method based on conserved quantities. In Section. 3, our two-stage PINN method based on conserved quantities is utilized to simulate abundant localized wave solutions including the one-soliton solution for the Boussinesq-Burgers equations and interaction solution for the classical Boussinesq-Burgers equations. Dynamic behaviors of soliton molecules for the Sawada-Kotera equation are also reproduced in Section. 4. In above two sections, given that we just use the original PINN in stage one, the performance of the two models can be evaluated in terms of the accuracy by comparing the results of the two stages. Then we present the relative $\mathbb{L}_2$ errors of these two methods and calculate error reduction rates. Finally, the conclusion and expectation are given in the last section.

\section{Methodology}
\subsection{The PINN method}
\quad

The physics-informed neural network method is briefly reviewed in this section \cite{Raissi2019}, which plays an important role in solving forward and inverse partial differential equations. We take the following (1+1)-dimensional nonlinear equation as an example to illustrate this method:
\begin{equation}\label{E1}
u_t+\mathcal{N}[u]=0, x \in\left[x_{0}, x_{1}\right], t \in\left[t_{0}, t_{1}\right],
\end{equation}
where $u=u(x,t)$ is the real-valued solution of this equation and $\mathcal{N}[\cdot]$ is a nonlinear differential operator in space. The governing equation $f(x,t)$ is defined by the left-hand-side of the Eq. \eqref{E1} above:
\begin{equation}\label{E2}
f:=u_t+\mathcal{N}[u].
\end{equation}

We aim to solve the initial-boundary value problem with the aid of physics-informed neural network technique. Meanwhile, the PINN method is introduced from three aspects as follows.

\textbf{(1) Structure establishment of PINN:}

Considering that the depth of neural network depends on the number of weighted layers, we construct a neural network of depth $L$ consisting of one input layer, $L-1$ hidden layers and one output layer. The $l$th ($l=0,1,\cdots,L$) layer has $N_l$ neurons, which represents that it transmits $N_l$-dimensional output vector $\mathbf{x}^l$ to the ($l+1$)th layer as the input data. The connection between layers is achieved by the following affine transformation $\mathcal{A}$ and activation function $\sigma(\cdot)$:
\begin{align}
\mathbf{x}^l=\sigma(\mathcal{A}_l(\mathbf{x}^{l-1}))=\sigma(\mathbf{w}^{l} \mathbf{x}^{l-1}+\mathbf{b}^{l}),	
\end{align}
where $\mathbf{w}^{l}\in \mathbb{R}^{N_{l} \times N_{l-1}}$ and $\mathbf{b}^{l}\in \mathbb{R}^{N_{l}}$ denote the weight matrix and bias vector of the $l$th layer, respectively. Thus, the relation between input $\mathbf{x}^0$ and output $u(\mathbf{x}^0,\boldsymbol{\Theta})$ is given by
\begin{align}
u(\mathbf{x}^0,\boldsymbol{\Theta})=(\mathcal{A}_L \circ \sigma \circ \mathcal{A}_{L-1} \circ \cdots \circ \sigma \circ \mathcal{A}_1)(\mathbf{x}^0),
\end{align}
and here $\boldsymbol{\Theta}=\left\{\mathbf{w}^{l}, \mathbf{b}^{l}\right\}_{l=1}^{L}$ represents the trainable parameters of PINN.

Before training a NN model, we need to initialize the parameters. Usually, the bias term is initialized to zero. There are many effective methods to initialize weight matrixes, such as Xavier initialization \cite{Glorot2010}, He initialization \cite{He2015}, etc. Given that the expression ability of the linear model is not enough, the activation function is used to add nonlinear factors to neural networks. The most frequently used nonlinear activation functions include $ReLU$ function, $Sigmoid$ function and $tanh$ function. In this paper, we select $tanh$ function as the activation function and initialize weights of the neural network with the Xavier initialization.

\textbf{(2) Parameter optimization of PINN:}

The essence of the training neural networks or deep learning models is to update the weights and biases. Based on the training data, our goal is to minimize the value of the loss function by optimizing the parameters of the neural network.

Assume we can obtain the initial-boundary dataset $\{x^i_u,t^i_u,u^i\}^{N_u}_{i=1}$ and the set of collocation points of $f(x,t)$, denoted by $\{x_{f}^i,t_{f}^i\}^{N_{f}}_{i=1}$. Then we construct the mean squared error function as the loss function to measure the difference between the predicted values and the true values of each iteration. The given information is investigated to merge into mean squared error, including the initial and boundary data as well as the governing equation:

\begin{equation}\label{E3}
MSE_1=MSE_u+MSE_f,
\end{equation}
where
\begin{equation}\label{E4}
MSE_u=\frac{1}{N_u}\sum^{N_u}_{i=1}|\Widehat{u}(x_u^i,t_u^i)-u^i|^2,
\end{equation}

\begin{equation}\label{E5}
MSE_{f}=\frac{1}{N_f}\sum^{N_f}_{i=1}|f(x_{f}^i,t_{f}^i)|^2.
\end{equation}
Here, $\{\Widehat{u}(x_u^i,t_u^i)\}^{N_u}_{i=1}$ denote the predicted results and the derivatives of the network $u$ with respect to time $t$ and space $x$ are derived by automatic differentiation \cite{AutomaticDifferentiation} to obtain $\{f(x_{f}^i,t_{f}^i)\}^{N_{f}}_{i=1}$. Based on MSE criteria, the parameters of neural networks are optimized to approach the initial and boundary training data and satisfy the structure imposed by \eqref{E1}. Several commonly used optimization methods of loss functions are: L-BFGS \cite{L-BFGS}, SGD, Adam and we apply L-BFGS method here. Hence, numerical solutions of the given domain and period can be obtained according to the trained PINN.

\textbf{(3) Capability Evaluation of PINN:}

Actually, the PINN method only involves above two aspects. However, in this paper, we aim to evaluate the performance of the PINN method in the circumstances of known solutions of Eq. \eqref{E1}.

We divide spatial region $[x_0,x_1]$ and time region $[t_0,t_1]$ into $N_x$ and  $N_t$ discrete equidistance points, respectively. Then the solution $u$ is discretized into $N_x \times N_t$ data points in the given spatiotemporal domain. We randomly select $N_u$ points of initial-boundary data on the above grids ($\mathcal{I} \cup \mathcal{B}, \mathcal{I}=[x_0+j\frac{x_1-x_0}{N_x-1},t_0],(j=0,1,\cdots,N_x-1), \mathcal{B}=[x, t_0+k\frac{t_1-t_0}{N_t-1}],(x=x_0\ {\rm or}\ x_1, k=0,1,\cdots,N_t-1)$) and obtain a random selection of $N_f$ collocation points of $f(x,t)$ in $\left[x_{0}, x_{1}\right] \times \left[t_{0}, t_{1}\right]$, which is not required to appear on grids. Thus, the training data in this case is $\{x^i_u,t^i_u,u^i\}^{N_u}_{i=1}$ and $\{x_{f}^i,t_{f}^i\}^{N_{f}}_{i=1}$. Given that the size of training data is only a small percentage of total data on grids, we calculate the relative $\mathbb{L}_2$ error ($RE$) of $N_x \times N_t$ data points on grids to evaluate the generalization ability of the PINN model:
\begin{align}
RE=\frac{\sqrt{\sum^{N_x-1}_{j=0} \sum^{N_t-1}_{k=0} |\Widehat{u}(x_0+j\frac{x_1-x_0}{N_x-1},t_0+k\frac{t_1-t_0}{N_t-1})-u^{j, k}|^2}}{\sqrt{\sum^{N_x-1}_{j=0} \sum^{N_t-1}_{k=0} |u^{j, k}|^2}},
\end{align}
where $\Widehat{u}(x_0+j\frac{x_1-x_0}{N_x-1},t_0+k\frac{t_1-t_0}{N_t-1})$ and $u^{j, k}$ represent the predictive value and true value, separately.

\subsection{Introduction of conserved quantities}
\quad

In this part, the introduction of conserved quantities is presented in brief \cite{LYS1999}.

For a finite-dimensional system, let $q_i, p_i(i=1,2,\cdots,n)$ be the generalized coordinates and momentums of the mechanical system. If Hamiltonian functions $H=H(q_i,p_i)$ exist, which satisfy
\begin{align}\label{E14}
\frac{d q_{i}}{d t}=\frac{\partial H}{\partial p_{i}}, \quad \frac{d p_{i}}{d t}=-\frac{\partial I}{\partial q_{i}}, \quad(i=1,2, \cdots, n)
\end{align}
then \eqref{E14} can be rewritten as
\begin{equation}\label{E15}
\begin{array}{c}
\dot{q}_{i}=\left\{q_{i}, H\right\}, \quad \dot{p}_{i}=\left\{p_{i}, H\right\}, \\
\dot{q}_{i}=\frac{d q_{i}}{d t}, \quad \dot{p}_{i}=\frac{d p_{i}}{d t},
\end{array}
\end{equation}
after introducing Poisson brackets
\begin{align}\label{E16}
\{F, G\}=\sum_{j=1}^{n}\left(\frac{\partial F}{\partial q_{j}} \frac{\partial G}{\partial p_{j}}-\frac{\partial F}{\partial p_{j}} \frac{\partial G}{\partial q_{j}}\right).
\end{align}
Besides, $q_i$ and $p_i$ satisfy the following relations
\begin{align}\label{E17}
\left\{q_{i}, q_{j}\right\}=\left\{p_{i}, p_{j}\right\}=0, \quad\left\{q_{i}, p_{j}\right\}=\delta_{i j}.	
\end{align}
Therefore, Eq. \eqref{E14} is called the Hamilton system.
If there is $I=I(q_i,p_i)$, which holds
\begin{align}\label{E18}
\frac{d I}{d t}=0,	
\end{align}
then $I$ is called a conserved quantity of Eq. \eqref{E14}.

For infinite dimensional systems, we take the following (1+1)-dimensional nonlinear equation as an example
\begin{align}\label{E19}
\Delta(x, t, u(x, t))=0,	
\end{align}
and then a conserved quantity $m_i$ can be defined similarly, which is time-independent and usually obtained by calculating the integral from $-\infty$ to $\infty$ with respect to a corresponding conserved density $\Gamma_{i}(x,t)$:
\begin{align}\label{E20}
m_i=\int_{-\infty}^{\infty} \Gamma_{i} {\rm dx},(i=1,2,\cdots).	
\end{align}
Then Eq. \eqref{E19} have the corresponding conservation law
\begin{align}\label{E21}
D_t \Gamma_{i}+ D_x	J_i=0,(i=1,2,\cdots)
\end{align}
which is satisfied for all solutions of \eqref{E19}. Here, $\Gamma_{i}(x,t)$ is the conserved density and $J_i(x,t)$ is the associated flux \cite{G1997}. The above formulas reveal the relationship between conserved quantities and conservation laws.

Integrable systems have infinite conserved quantities, which is a pretty significant property. Generally speaking, it's not plain to derive conserved quantities. Sometimes, the first few conserved quantities in physical problems usually correspond to the conservation of mass, momentum, or energy. Others may facilitate the research of the quantitative and qualitative properties of solutions.

\subsection{The two-stage PINN method based on conserved quantities.}
\quad

The main purpose of this article is to put forward a more targeted PINN algorithm of nonlinear mathematical physics. We try to introduce more features of integrable systems into neural networks to improve the precision and reliability. This part epitomizes the main idea of the two-stage PINN method based on conserved quantities.

\textbf{(1) Stage One:}

In the first stage, we use the original PINN method which is mentioned in section 2.1. Under the principle of minimizing the mean squared error loss, we can acquire the numerical solution $\Widehat{u}_1(x,t)$ of the given domain and period after parameter optimization.

\textbf{(2) Stage Two:}

On the basis of stage one, we make the following improvements. Based on conserved quantities, we aim to achieve further optimization of the numerical solution $\Widehat{u}_1(x,t)$ in the first stage.

Firstly, we should gain a conserved quantity $m(t)$ of the corresponding equation, which evidently depends on  the choice of the solution $u$ and is actually time-independent, i.e. $\frac{d m(t)}{d t}=0$. Based on the initial data of $u(x,t)$, the conserved quantity $m(t_0)$ can be calculated and taken as the criterion.  We randomly select $N_c$ different moments and measure the corresponding conserved quantities $\{m(t_m^i)\}^{N_c}_{i=1}$. Our goal is to make $\{m(t_m^i)\}^{N_c}_{i=1}$ approach the theoretical value $m(t_0)$ as close as possible.

According to the above analysis, the mean squared error loss of the original PINN is changed into:
\begin{equation}\label{E6}
MSE_2=MSE_u+MSE_f+MSE_s+MSE_m,
\end{equation}
where
\begin{equation}\label{E7}
MSE_s=\frac{1}{N_s}\sum^{N_s}_{i=1}|\Widehat{u}(x_s^i,t_s^i)-\Widehat{u}_1(x_s^i,t_s^i)|^2,	
\end{equation}
\begin{equation}\label{E8}
MSE_m=\frac{1}{N_c}\sum^{N_c}_{i=1}|m(t_m^i)-m(t_0)|^2.	
\end{equation}
Here, $\Widehat{u}(x,t)$ denotes the numerical solution obtained in stage two and $MSE_s$ measures the difference of  numerical results between two stages at $\{x^i_s,t^i_s\}^{N_s}_{i=1}$, which implies that further optimization is based on stage one and $N_s$ points $\{x^i_s,t^i_s,\Widehat{u}_1(x_s^i,t_s^i),\Widehat{u}(x_s^i,t_s^i)\}^{N_s}_{i=1}$ are sampled randomly on the grids. Meanwhile, $MSE_m$ reflects the constraint of the conserved quantity.

With regard to the calculation of conserved quantities, we adopt the method of numerical integral by using summation instead of integrals. Suppose $\mathcal{M}[u]$ is a conserved density ($\mathcal{M}[\cdot]$ denotes a differential operator) and we divide spatial region $[x_0,x_1]$ into $N_x$ discrete equidistance points with time region $[t_0,t_1]$ into $N_t$ discrete equidistance points, then $\mathcal{M}[u]$ is discretized into $N_x \times N_t$ data points and the formulas of $m(t_0)$ and $m(t_m^i)$ are derived:
\begin{equation}\label{E9}
m(t_0)=\int_{x_0}^{x_1} \mathcal{M}[u](x, t_0) {\rm dx} \approx \sum^{N_x}_{j=2} \mathcal{M}[u](x^j, t_0) \frac{x_1-x_0}{N_x-1},
\end{equation}
\begin{equation}\label{E10}
m(t_m^i)=\int_{x_0}^{x_1} \mathcal{M}[\Widehat{u}](x, t_m^i) {\rm dx} \approx \sum^{N_x}_{j=2} \mathcal{M}[\Widehat{u}](x^j, t_m^i) \frac{x_1-x_0}{N_x-1},	
\end{equation}
where $\mathcal{M}[u](x^j, t_0)$ and $\mathcal{M}[\Widehat{u}](x^j, t_m^i)$ represent the true value and predictive value, respectively.

In the original PINN method, the loss $MSE_u$ and $MSE_f$ reflect the local constraints at certain points solely, which are selected stochastically. However, in stage two, the calculation of conserved quantities involves the integral operation. It is widely known that at any given time, conserved quantities mirror the global property of the solution $u$ in $\left[x_{0}, x_{1}\right]$. Thus, our practice to introduce the measurement of this global property into the mean squared error loss is meaningful and is a kind of method to impose constraints from a global perspective.

Similarly, if we consider $k$ conserved quantities $\emph{\textbf{m}}=(m_1,m_2,\cdots,m_k)$, $MSE_m$ is transformed into:
\begin{equation}\label{E11}
MSE_{\emph{\textbf{m}}}=\frac{1}{N_c}\sum^{k}_{j=1}\sum^{N_c}_{i=1}|m_j(t_m^i)-m_j(t_0)|^2.	
\end{equation}

\begin{figure}
\centering
\includegraphics[width=6cm,height=4cm]{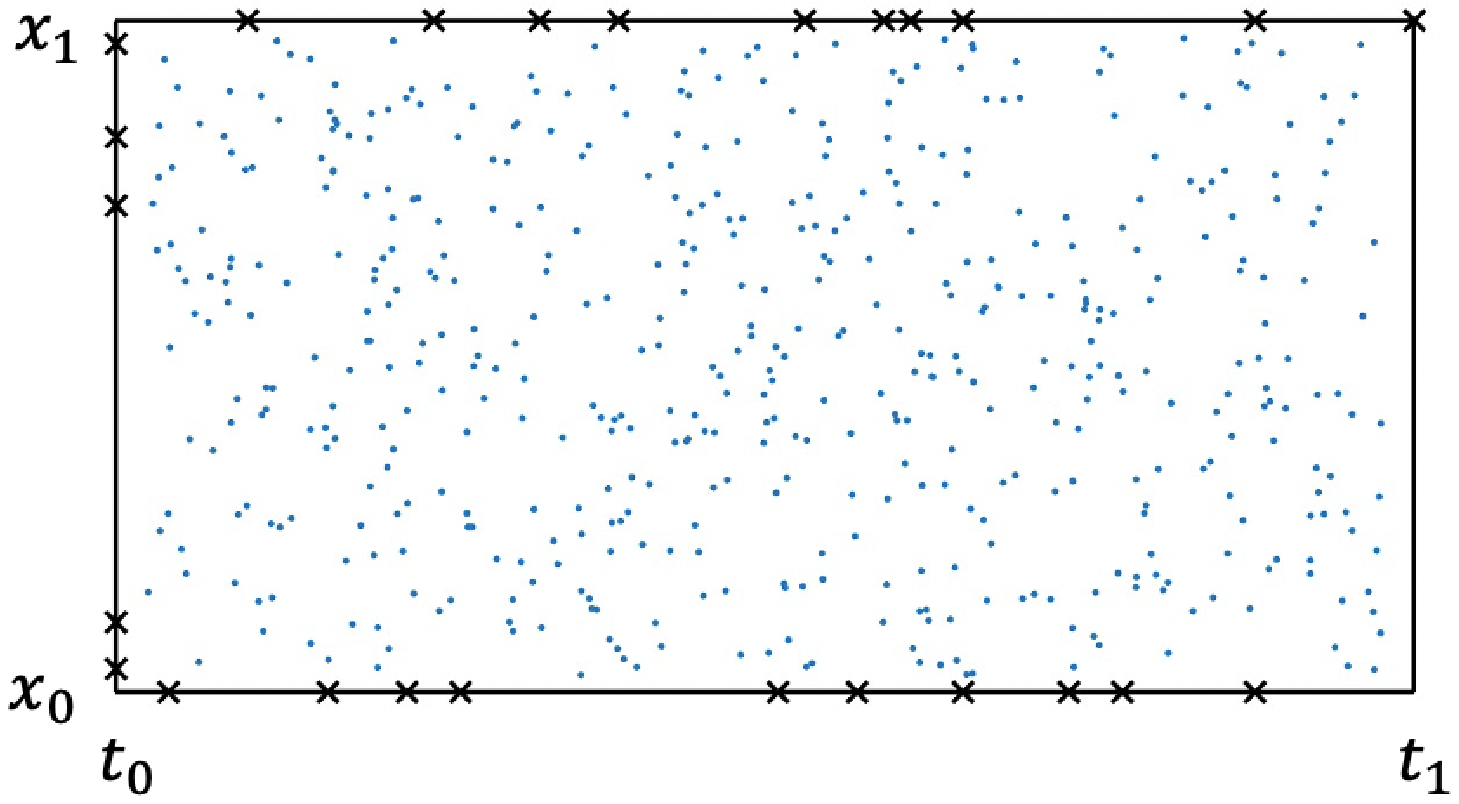}
$a$
\includegraphics[width=6cm,height=4cm]{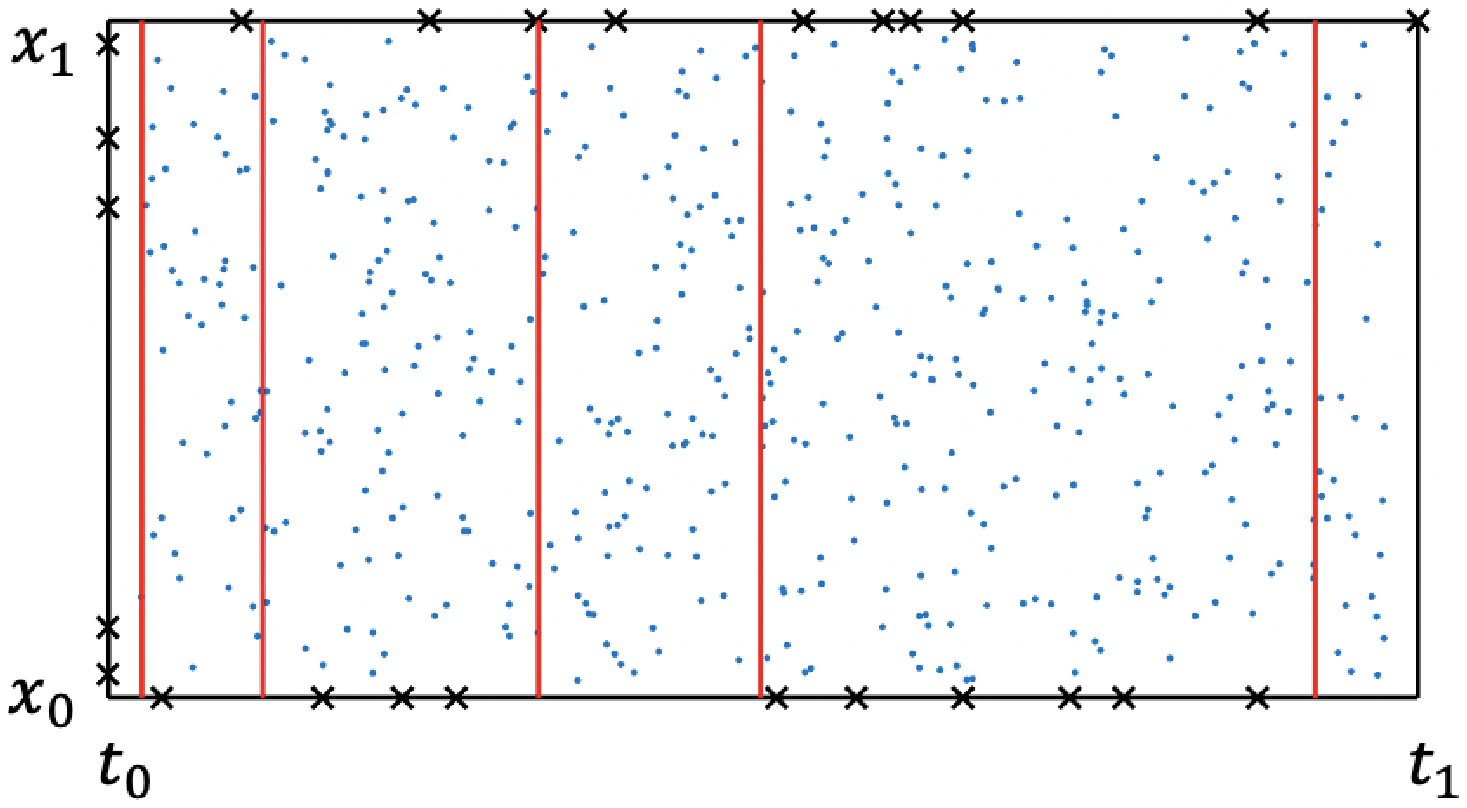}
$b$
\caption{(Color online) Schematic diagrams of constraints: (a) Constraints of the original PINN method;
(b) Constraints of the two-stage PINN method based on conserved quantities.}
\end{figure}

With regard to two methods above, we display the schematic diagrams of constraints in Fig. 1. Black crosses imply that these selected points need to meet initial-boundary conditions and blue dots represent the random selection of points should satisfy the structure imposed by the governing equation. They are both local constraints. The two-stage PINN method based on conserved quantities differs from the original PINN method in that it takes conserved quantities into consideration to impose constraints globally. We use the red lines to denote calculation of conserved quantity at certain moments, which involves the integral operation and the method of numerical integral is adopted by using summation instead of integrals.

Moreover, Fig. 2 shows a sketch of the two-stage PINN method based on conserved quantities, where $maxit$ denotes the maximum number of iterations.
\begin{figure}
\centering
\includegraphics[width=15cm,height=15cm]{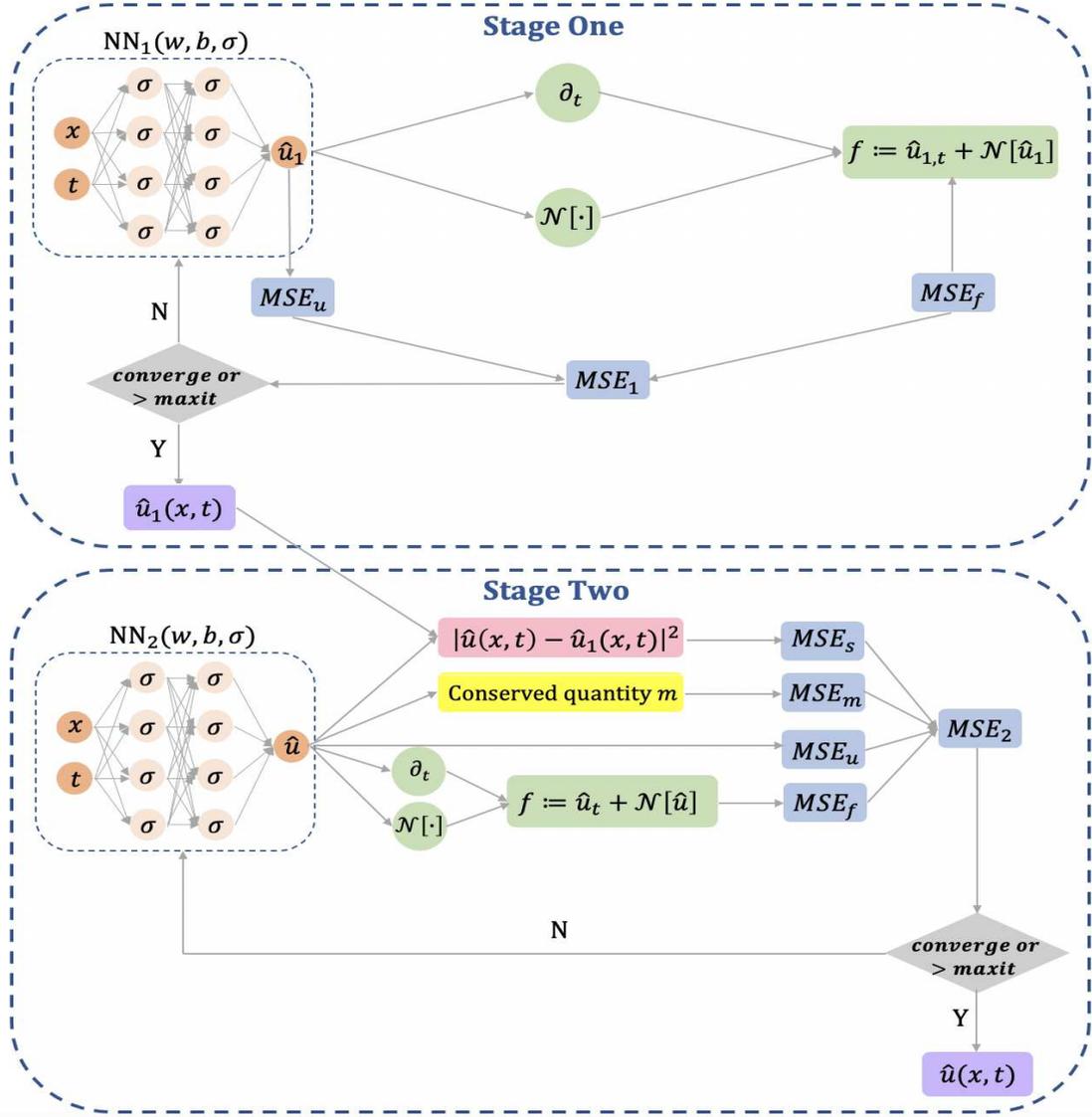}
\caption{(Color online) Schematic diagram of the two-stage physics-informed neural network method based on conserved quantities.}
\end{figure}

This method is established on the assumption that heights of background waves in the region $\mathcal{D}$ ($\mathcal{D}=[x,t], x \in\left(-\infty,x_{0}\right] \bigcup \left[x_{1},+\infty\right), t \in\left[t_{0}, t_{1}\right]$) are almost consistent. Therefore, we can integral from $x_0$ to $x_1$ with respect to a conserved density to represent conserved quantities instead of from $-\infty$ to $+\infty$. It is a reasonable assumption in the sense that it can be easily satisfied by localized wave solutions, which are widely considered in the field of integrable systems to describe various physical phenomena.

Actually, a natural idea is to take the following formula
\begin{align}\label{E12}
MSE=MSE_u+MSE_f+MSE_m,	
\end{align}
as the optimization objective of PINN directly rather than carrying out this two-stage method. However, the result was a disappointment and even worse than the original PINN method. After the analysis, we are of the opinion that the optimization is dominated by $MSE_m$, and finally it converges to other local optimal point which causes unsatisfactory results. Consequently, we propose the two-stage PINN method to improve it. This method not only considers the global property measured by conserved quantities, but also further optimizes parameters of the original PINN. The numerical results also show that it can avoid converging to other non-ideal local optimums.

All codes are based upon Python 3.7 and Tensorflow 1.15, and the presented numerical experiments are run on a MacBook Pro computer with 2.3 GHz Intel Core i5 processor and 16-GB memory.

\section{Data-driven one-soliton solution of the Boussinesq-Burgers equations and interaction solution of the classical Boussinesq-Burgers equations}

In this section, we will apply the two-stage PINN method to numerically solve integrable equations and then contrast the simulation results of the two models: the PINN and two-stage PINN based on conserved quantities. Considering that we just use the original PINN in stage one, the performance of the two models can be evaluated in terms of the accuracy by comparing the results of the two stages.

The current research of coupled equations with the aid of neural networks is relatively less than that of the single equation and thus we mainly consider the coupled equations here: the Boussinesq-Burgers equations \cite{Kaup1975, Kupershmidt1985} and the classical Boussinesq-Burgers equations \cite{Date1978,Geng1999, Ito1984, Kawamoto1984, Gu1990}.
\subsection{One-soliton solution of the Boussinesq-Burgers equations}

Here, we investigate the Boussinesq-Burgers equations \cite{Kaup1975, Kupershmidt1985} with the first kind of boundary condition (Dirichlet boundary condition)
\begin{equation}\label{eq3.1}
\begin{split}
\begin{cases}
u_{t}+2 u u_{x}-\frac{1}{2} v_{x}=0,\\
v_{t}+2(u v)_{x}-\frac{1}{2} u_{x x x}=0, x\in[x_0,x_1],t\in[t_0,t_1],\\
u(x,t_0)=u_0(x),\\
v(x,t_0)=v_0(x),\\
u(x_0,t)=a_1(t), u(x_1,t)=a_2(t),\\
v(x_0,t)=a_3(t), v(x_1,t)=a_4(t).
\end{cases}
\end{split}
\end{equation}
Wang et al. \cite{Wang2011} studied the Lax pair, B\"acklund transformation and multi-soliton solutions for the Boussinesq-Burgers equations. Many researchers also obtained a variety of soliton solutions and some exact interaction solutions of the Boussinesq-Burgers equations, which describe the propagation of shallow water waves \cite{Wazwaz2017, Wang2014}. In Ref. \cite{Rady2010}, Rady et al. have derived the multi-soliton solution of this equation. Firstly, they consider the following function transformation
\begin{align}\label{eq3.2}
v=\lambda u_{x}+\beta,	
\end{align}
and set $\lambda=-1, \beta=0$. In the light of the idea of homogeneous balance method \cite{Wang1996} as well as the B\"acklund transformation, the multi-soliton solution can be obtained
\begin{align}\label{eq3.3}
&u=\frac{1}{2} \frac{\sum_{i=1}^{n} k_{i} \exp \left(k_{i}\left(x-2\left(a+\frac{k_{i}}{4}\right) t\right)\right)}{1+\sum_{i=1}^{n} \exp \left(k_{i}\left(x-2\left(a+\frac{k_{i}}{4}\right) t\right)\right)}+a,\nonumber\\
&v=-u_x.
\end{align}

In this case, the governing equations $f_1(x,t)$ and $f_2(x,t)$ are as follows
\begin{align}\label{eq3.4}
&f_1:=u_{t}+2 u u_{x}-\frac{1}{2} v_{x},\nonumber\\
&f_2:=v_{t}+2(u v)_{x}-\frac{1}{2} u_{x x x}.
\end{align}

When $n=1$, the corresponding initial-boundary conditions are given by

\begin{align}\label{eq3.5}
&u(-20,t)=-\frac{e^{20+\frac{7t}{2}}}{2(1+e^{20+\frac{7t}{2}})}+2, u(20,t)=-\frac{e^{-20+\frac{7t}{2}}}{2(1+e^{-20+\frac{7t}{2}})}+2,\nonumber\\
&v(-20,t)=-\frac{\mathrm{e}^{20+\frac{7 t}{2}}}{2\left(1+\mathrm{e}^{20+\frac{7 t}{2}}\right)}+\frac{\left(\mathrm{e}^{20+\frac{7 t}{2}}\right)^{2}}{2\left(1+\mathrm{e}^{20+\frac{7 t}{2}}\right)^{2}}, v(20,t)=-\frac{\mathrm{e}^{-20+\frac{7 t}{2}}}{2\left(1+\mathrm{e}^{-20+\frac{7 t}{2}}\right)}+\frac{\left(\mathrm{e}^{-20+\frac{7 t}{2}}\right)^{2}}{2\left(1+\mathrm{e}^{-20+\frac{7 t}{2}}\right)^{2}},\nonumber\\
&u_0(x)=-\frac{e^{-x-7}}{2(1+e^{-x-7})}+2, v_0(x)=-\frac{e^{-x-7}}{2(1+e^{-x-7})}+\frac{(e^{-x-7})^2}{2(1+e^{-x-7})^2},
\end{align}
after choosing corresponding parameters as $a=2, k_1=-1, [x_0,x_1]=[-20,20],  [t_0,t_1]=[-2,2]$. To obtain the training data,  we divide the spatial region $[x_0,x_1]=[-20,20]$ and time region $[t_0,t_1]=[-2,2]$ into $N_x=1025$ and $N_t=201$ discrete equidistance points, separately. Thus, the solutions $u$ and $v$ are both discretized into $1025 \times 201$ data points in the given spatiotemporal domain. We randomly select $N_u=100$ points from the initial-boundary dataset and proceed by sampling $N_f=10000$ collocation points via the Latin hypercube sampling method \cite{Stein1987}. A 8-layer feedforward neural network with 40 neurons per hidden layer is constructed to learn the one-soliton solution of the Boussinesq-Burgers equations. In addition, we use the hyperbolic tangent ($tanh$) activation function and initialize weights of the neural network with the Xavier initialization. The derivatives of the network $u, v$ with respect to time $t$ and space $x$ are derived by automatic differentiation.

We utilize the L-BFGS algorithm to optimize loss functions. Obviously, $v$ is a conserved density of the Boussinesq-Burgers equations and we select $m$ defined by
\begin{align}\label{eq3.6}
m=\int_{x_0}^{x_1} v {\rm dx} \approx \sum^{N_x}_{j=2} v(x^j, t) \frac{x_1-x_0}{N_x-1},
\end{align}
as the conserved quantity adopted in two-stage PINN. The loss function of stage one is \eqref{E3} and that of stage two is \eqref{E6} where we choose $N_s=10000, N_c=20$ and $MSE_m$ is given by
\begin{align}\label{eq3.7}
MSE_m&=\frac{1}{N_c}\sum^{N_c}_{i=1}|m(t_m^i)-m(t_0)|^2	\nonumber\\
&\approx \frac{1}{N_c}\sum^{N_c}_{i=1}\bigg|\sum^{N_x}_{j=2} \Widehat{v}(x^j, t_m^i) \frac{x_1-x_0}{N_x-1}-\sum^{N_x}_{j=2} v(x^j, t_0) \frac{x_1-x_0}{N_x-1}\bigg|^2,
\end{align}
where $v(x^j, t_0)$ and $\Widehat{v}(x^j, t_m^i)$ represent the true value and predictive value, respectively.

Ultimately, the data-driven one-soliton solution of the Boussinesq-Burgers equations is obtained by two-stage PINN method based on conserved quantities.

Fig. 3 displays the density diagrams of the one-soliton solution, comparison between the predicted solutions and exact solutions as well as the error density diagrams. In the bottom panel of Fig. 3 (a) and Fig. 3 (c), we show the comparison between exact solutions and predicted solutions at different time points $t=-1.5, 0, 1.5$. Obviously, both $u$ and $v$ propagate along the positive direction of the $x$-axis as time goes by. Through contrastive analysis, one-soliton solution can be successfully simulated by two-stage PINN method with high accuracy. In Fig. 4, the three-dimensional plots of predicted one-soliton solutions $u(x,t)$ and $v(x,t)$ are showed respectively, where $v(x,t)$ is a dark soliton solution.

\begin{figure}
\centering
\includegraphics[width=7cm,height=5cm]{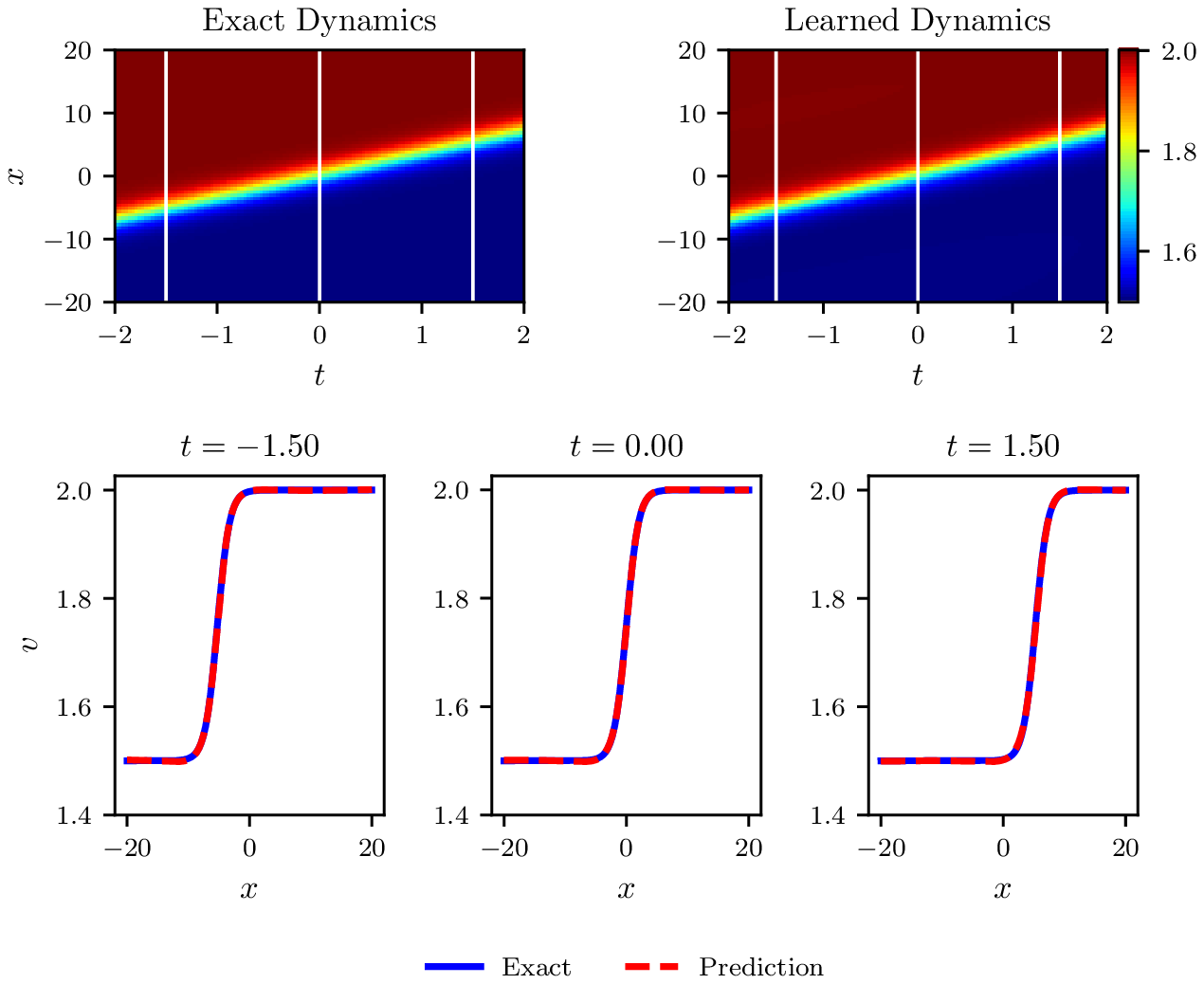}
$a$
\includegraphics[width=7cm,height=5cm]{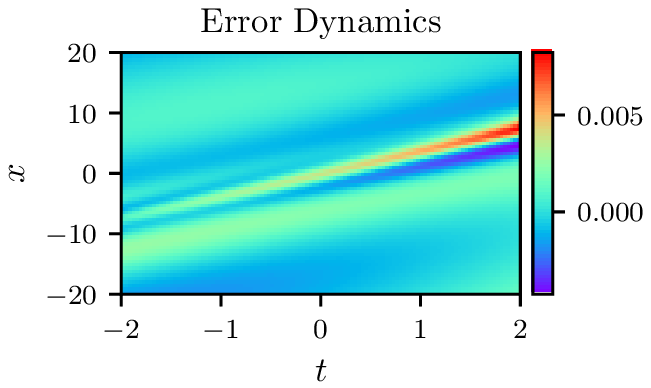}
$b$\\
\includegraphics[width=7cm,height=5cm]{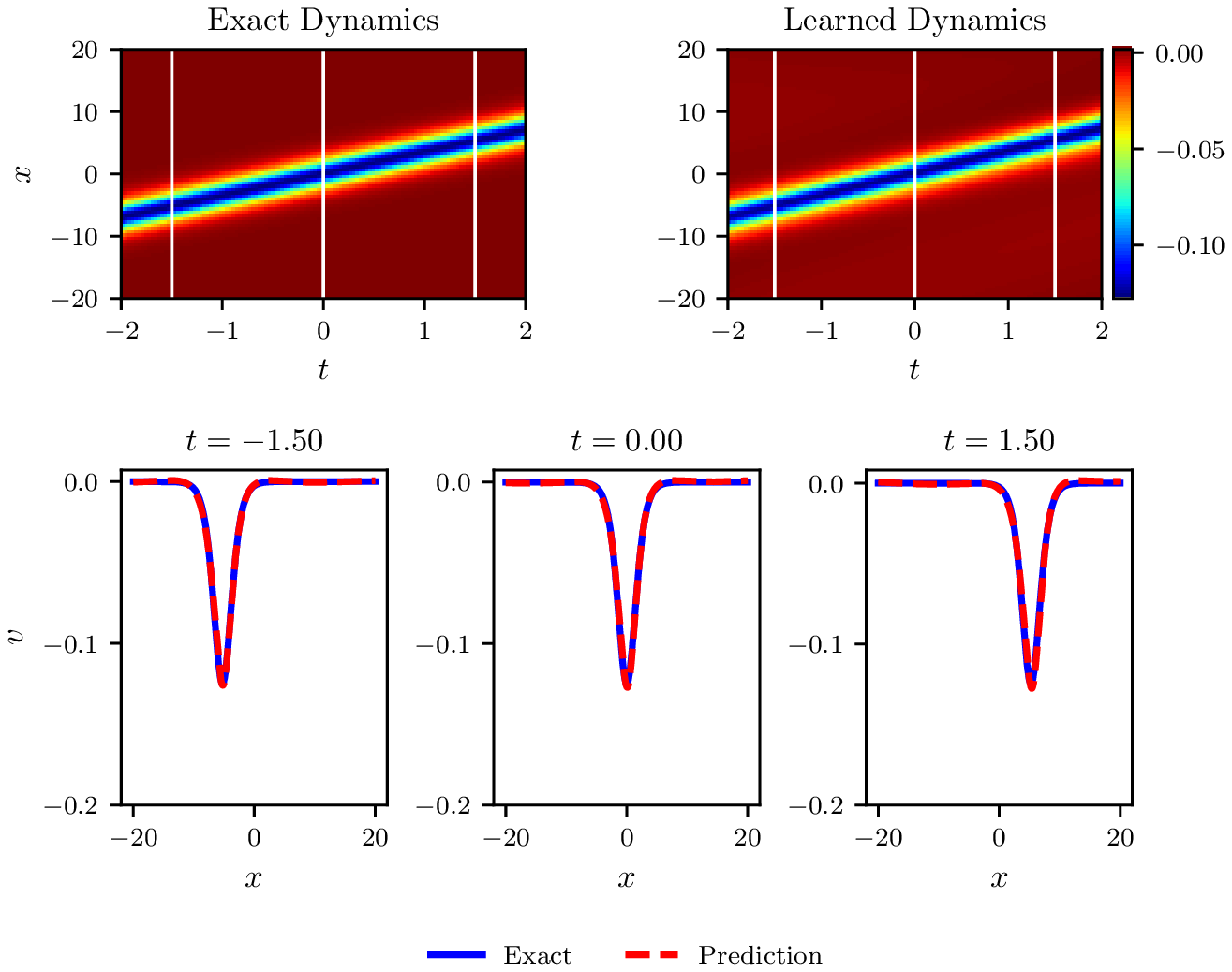}
$c$
\includegraphics[width=7cm,height=5cm]{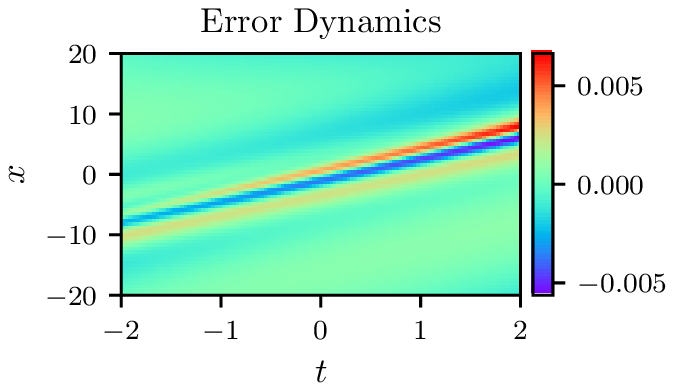}
$d$
\caption{(Color online) One-soliton solution $u(x,t)$ and $v(x,t)$ of the Boussinesq-Burgers equations by two-stage PINN based on conserved quantities: (a) The density diagrams and comparison between the predicted solutions and exact solutions at the three temporal snapshots of $u(x,t)$; (b) The error density diagram of $u(x,t)$; (c) The density diagrams and comparison between the predicted solutions and exact solutions at the three temporal snapshots of $v(x,t)$; (d) The error density diagram of $v(x,t)$.}
\end{figure}

\begin{figure}
\centering
\includegraphics[width=6.5cm,height=5cm]{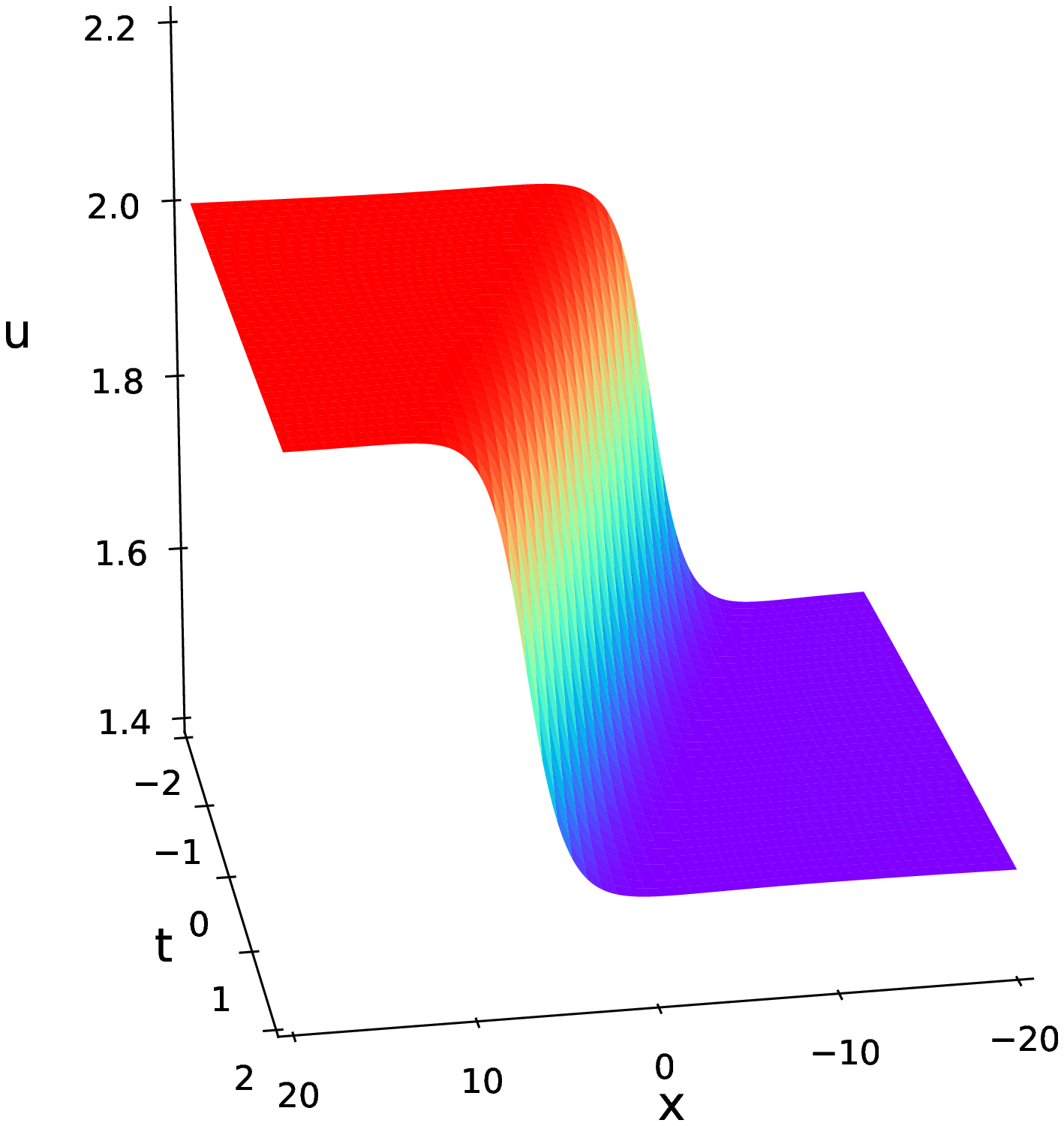}
$a$
\includegraphics[width=6.5cm,height=5cm]{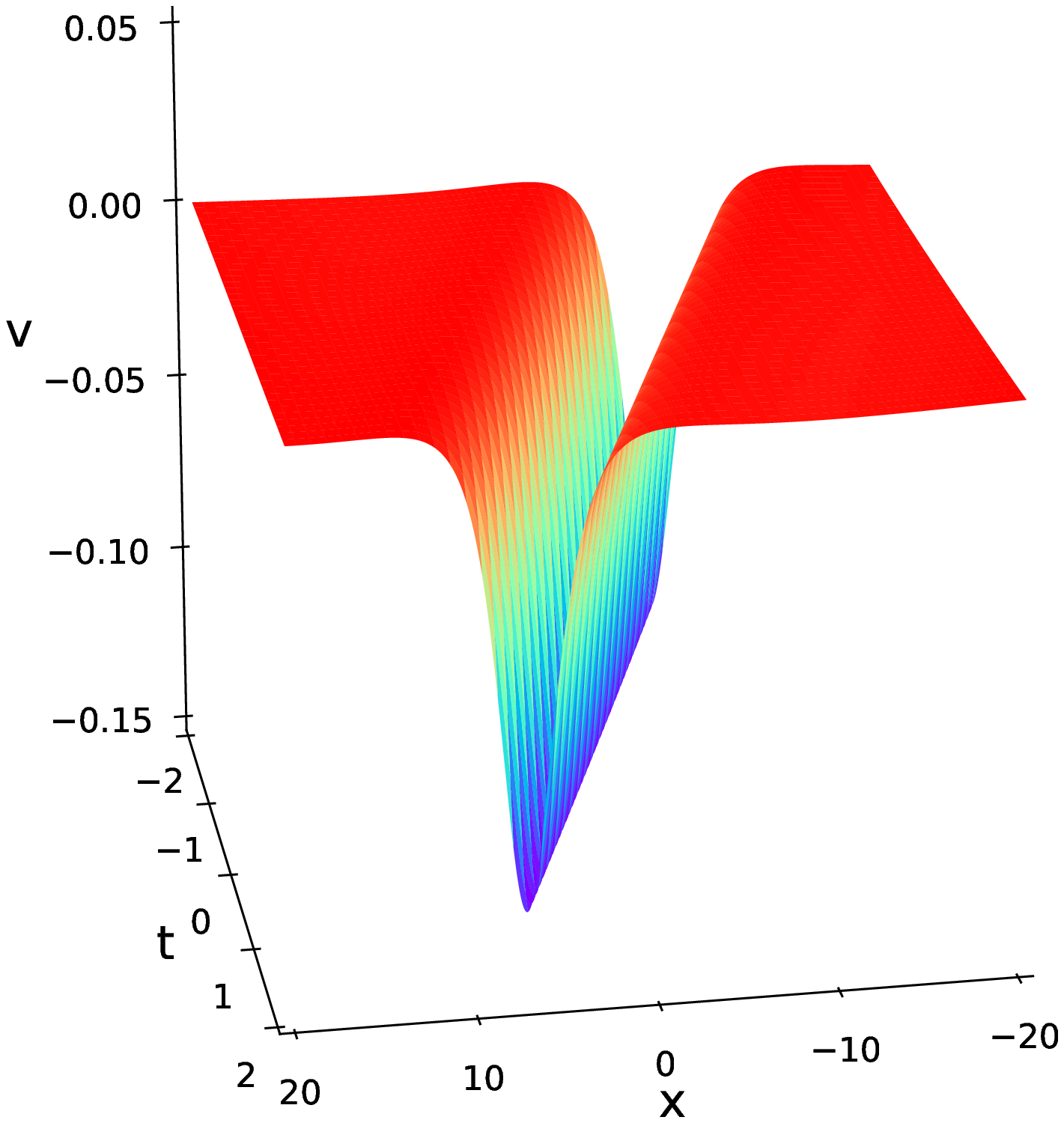}
$b$
\caption{(Color online) One-soliton solution $u(x,t)$ and $v(x,t)$ of the Boussinesq-Burgers equations by two-stage PINN based on conserved quantities: (a) The three-dimensional plot of $u(x,t)$;
(b) The three-dimensional plot of $v(x,t)$.}
\end{figure}

In stage one, the original PINN is applied. After 168 times iterations in about 41.7379 seconds, the relative $\mathbb{L}_2$ error of $u$ is 8.965473e-04 and that of $v$ is 4.750580e-02. In stage two, where the conserved quantity is considered, the relative $\mathbb{L}_2$ error of $u$ is 7.343612e-04 and that of $v$ is 3.776971e-02 after 1382 times iterations in about 407.5796 seconds. To compare the performance of two methods, error reduction rate ($ERR$) can be obtained according to the relative $\mathbb{L}_2$ error of PINN method ($RE_1$) and that of two-stage PINN method based on conserved quantities ($RE_2$):
\begin{align}\label{eq3.8}
ERR=\frac{RE_1-RE_2}{RE_1}.	
\end{align}
By calculation, the error reduction rate ($ERR$) of $u$ is $18.09\%$ and that of $v$ is $20.49\%$, which are presented in Table. 1. It turns out that our proposed two-stage PINN method based on conserved quantities can improve prediction accuracy and gain better generalization.

\begin{table}[htbp]
  \caption{One-soliton solution of the Boussinesq-Burgers equations: relative $\mathbb{L}_2$ errors of PINN and two-stage PINN based on conserved quantities as well as error reduction rates.}

  \centering
  \begin{tabular}{p{3.38cm}|p{3cm}p{4cm}p{4cm}}
  \toprule
  \textbf{\diagbox[height=0.7cm]{Solution}{Method}} &\textbf{\quad PINN} &\textbf{Two-stage PINN} &\textbf{Error reduction rate} \\
  \midrule
  \textbf{\quad\quad\quad\quad u}   &8.965473e-04&\quad\quad7.343612e-04&\quad\quad\quad18.09\%\\
  \textbf{\quad\quad\quad\quad v}   &4.750580e-02&\quad\quad3.776971e-02&\quad\quad\quad20.49\%\\
  \bottomrule
  \end{tabular}
\end{table}

\subsection{Interaction solution of the classical Boussinesq-Burgers equations}
\quad

In this part, we consider the classical Boussinesq-Burgers (CBB) equations \cite{Date1978,Geng1999, Ito1984, Kawamoto1984, Gu1990}
\begin{align}\label{eq3.9}
&u_{t}=\frac{1}{2}(\beta-1) u_{x x}+2 u u_{x}+\frac{1}{2} v_{x}, \nonumber\\
&v_{t}=\beta\left(1-\frac{\beta}{2}\right) u_{x x x}+\frac{1}{2}(1-\beta) v_{x x}+2(u v)_{x},	
\end{align}
where $u=u(x,t)$ and $v=v(x,t)$ are real-valued solutions and $\beta$ is an arbitrary constant. Obviously, the classical Boussinesq-Burgers equations are equivalent to the Boussinesq-Burgers equations under the condition $(u, v, x, t, \beta) \rightarrow(-u,-v,-x,-t, 1)$.  Moreover, Darboux transformations and soliton solutions of the classical Boussinesq-Burgers equations have been given in Ref. \cite{Xu2007}. Some scholars also have studied the finite-band solutions \cite{Geng1999}, rational solutions \cite{LiHuWu2018}, conservation laws and dynamical behaviors \cite{Jiang2019}. Dong et al. \cite{Dong2019} applied the consistent tanh expansion (CTE) to study the interaction solution for this equation given by
\begin{align}\label{eq3.10}
&u=u_0+u_1 \tanh(w),\nonumber\\
&v=v_0+v_1 \tanh(w)+v_2 \tanh(w)^2,
\end{align}
where
\begin{align}\label{eq3.11}
&u_1=\frac{w_x}{2},\quad u_0=\frac{2 w_t-w_{xx}}{4 w_x},\nonumber\\
&v_2=\frac{\beta w_x^2}{2}-w_x^2,\quad v_1=w_{xx}-\frac{\beta w_{xx}}{2},\nonumber\\
&v_0=-\frac{(\beta-2)(2 w_x^4-w_x w_{xxx}+2 w_x w_{xt}+w_{xx}^2-2 w_{xx} w_t)}{4 w_x^2},
\end{align}
and the interaction between soliton and resonance has the following form
\begin{align}\label{eq3.12}
&w=px+qt+\frac{1}{2} \ln \left(1+\sum_{i=1}^{n} \exp \left(p_{i} x+q_{i} t\right)\right), i=1,2, \ldots \nonumber\\
&q_{i}=\frac{p_{i}\left(2 q+p_{i} p+2 p^{2}\right)}{2 p}, \quad i=1,2, \ldots	
\end{align}
Here, we select the parameters as follows:
\begin{align}\label{eq3.13}
n=1,\quad p=1,\quad q=1,\quad \beta=1,\quad p_1=2.	
\end{align}

We focus on the classical Boussinesq-Burgers (CBB) equations with the first kind of boundary condition (Dirichlet boundary condition)
\begin{equation}\label{eq3.14}
\begin{split}
\begin{cases}
u_{t}=2 u u_{x}+\frac{1}{2} v_{x},\\
v_{t}=\frac{1}{2} u_{x x x}+2(u v)_{x}, x\in[x_0,x_1],t\in[t_0,t_1],\\
u(x,t_0)=u_0(x),\\
v(x,t_0)=v_0(x),\\
u(x_0,t)=a_1(t), u(x_1,t)=a_2(t),\\
v(x_0,t)=a_3(t), v(x_1,t)=a_4(t).
\end{cases}
\end{split}
\end{equation}
After setting $[x_0,x_1]=[-10,15], [t_0,t_1]=[-3,2]$, initial conditions of the  interaction solution above are obtained as follows

\begin{align}\label{eq3.15}
&u_0(x)=\frac{(4 e^{-36+4x}+4 e^{-18+2x}+1)(\tanh(x-3+\frac{\ln(1+e^{-18+2x})}{2})+1)}{4(e^{-18+2x})^2+6 e^{-18+2x}+2},\nonumber\\
&v_0(x)=\frac{2 e^{-18+2x} \sinh(x-3+\frac{\ln(1+e^{-18+2x})}{2}) \cosh(x-3+\frac{\ln(1+e^{-18+2x})}{2})+2 e^{-18+2x} \cosh^2(x-3+\frac{\ln(1+e^{-18+2x})}{2})}{2 (1+e^{-18+2x})^2 \cosh^2(x-3+\frac{\ln(1+e^{-18+2x})}{2})}\nonumber\\
&\quad\quad\quad+\frac{4 e^{-36+4x}+4 e^{-18+2x}+1}{2 (1+e^{-18+2x})^2 \cosh^2(x-3+\frac{\ln(1+e^{-18+2x})}{2})}.
\end{align}
as well as corresponding boundary conditions, which are no longer presented here due to space limitation.

We construct a 9-layer feedforward neural network with 40 neurons per hidden layer to learn the interaction solution between a soliton and one resonant. With the help of MATLAB, spatial region $[x_0,x_1]=[-10,15]$ and time region $[t_0,t_1]=[-3,2]$ are divided into $N_x=1025$ and  $N_t=201$ discrete equidistance points, respectively. After adopting the same generation and sampling method of training data in Section 3.1, we randomly select $N_u=100$ points from the initial-boundary dataset and $N_f=10000$ collocation points.

Here, $v$ is a conserved density of the classical Boussinesq-Burgers equations and the conserved quantity adopted in two-stage PINN is defined just as \eqref{eq3.6}. The loss function of stage one is \eqref{E3} and that of stage two is \eqref{E6} where we choose $N_s=10000, N_c=20$ and the computing formula of $MSE_m$ is consistent with \eqref{eq3.7}. In addition, the L-BFGS algorithm to optimize loss functions is the same in Section 3.1, as well as the Xavier initialization and the hyperbolic tangent ($tanh$) activation function.

The two-stage PINN method based on conserved quantities
 eventually succeed in numerical simulations of the interaction solution between a soliton and one resonant.

In Fig. 5, the density diagrams of interaction solution, comparison between the predicted solutions and exact solutions as well as the error density diagrams are plotted. Form the bottom panel of Fig. 5 (a) and Fig. 5 (c), it implies there is little difference between exact solutions and predicted solutions. Meanwhile, it can be seen that two peaks converge into one of higher amplitude according to the wave propagation pattern of $v(x,t)$ and it propagates along the negative direction of the $x$-axis. Fig. 6 displays the predicted 3D plots of the interaction solution, which show interaction behaviors between soliton and resonance.

\begin{figure}
\centering
\includegraphics[width=7cm,height=5cm]{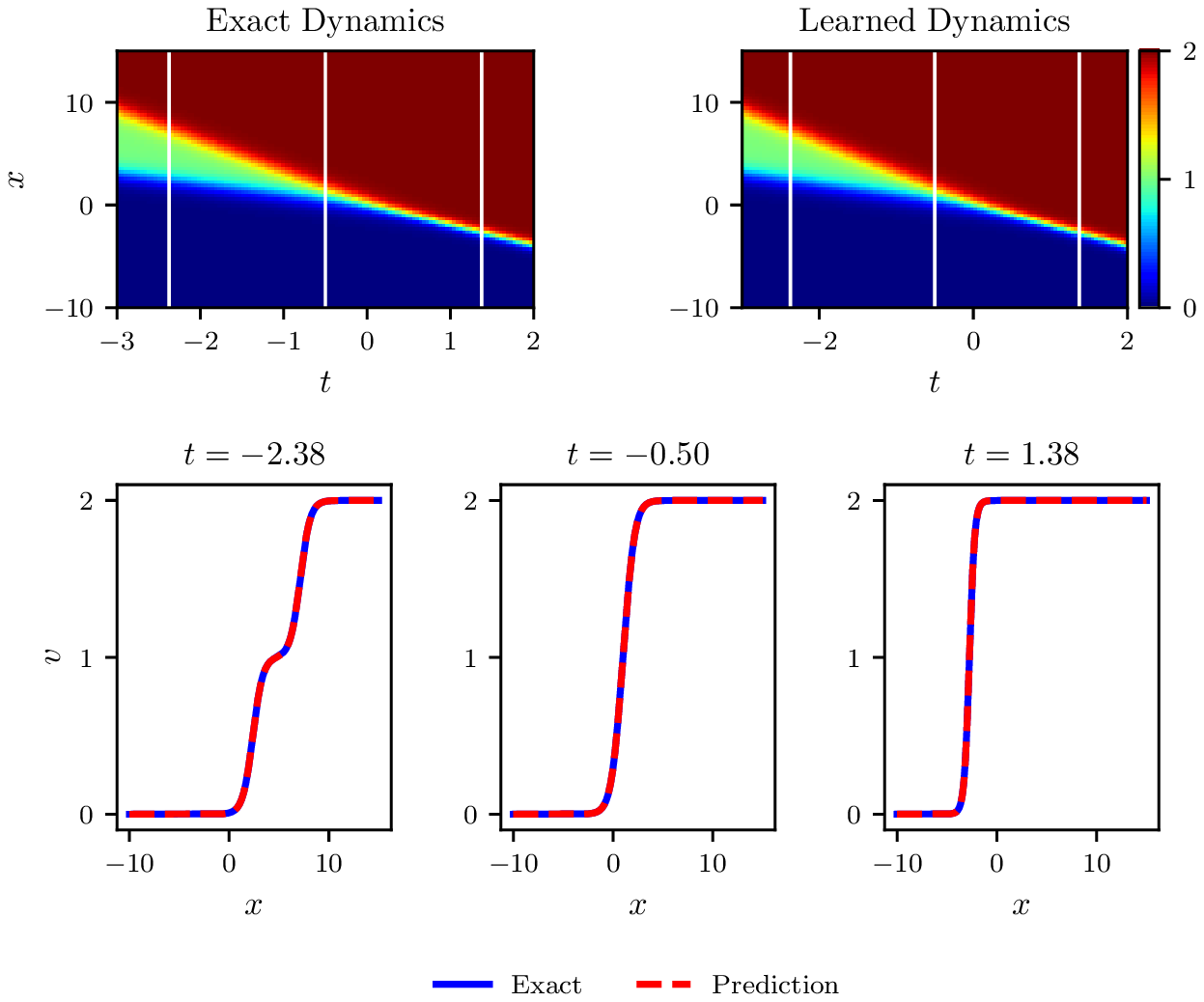}
$a$
\includegraphics[width=7cm,height=5cm]{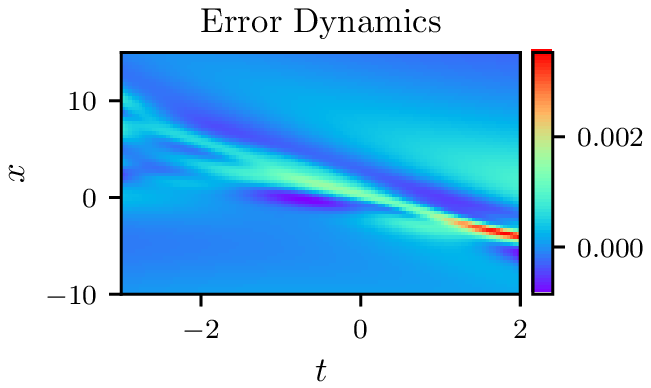}
$b$\\
\includegraphics[width=7cm,height=5cm]{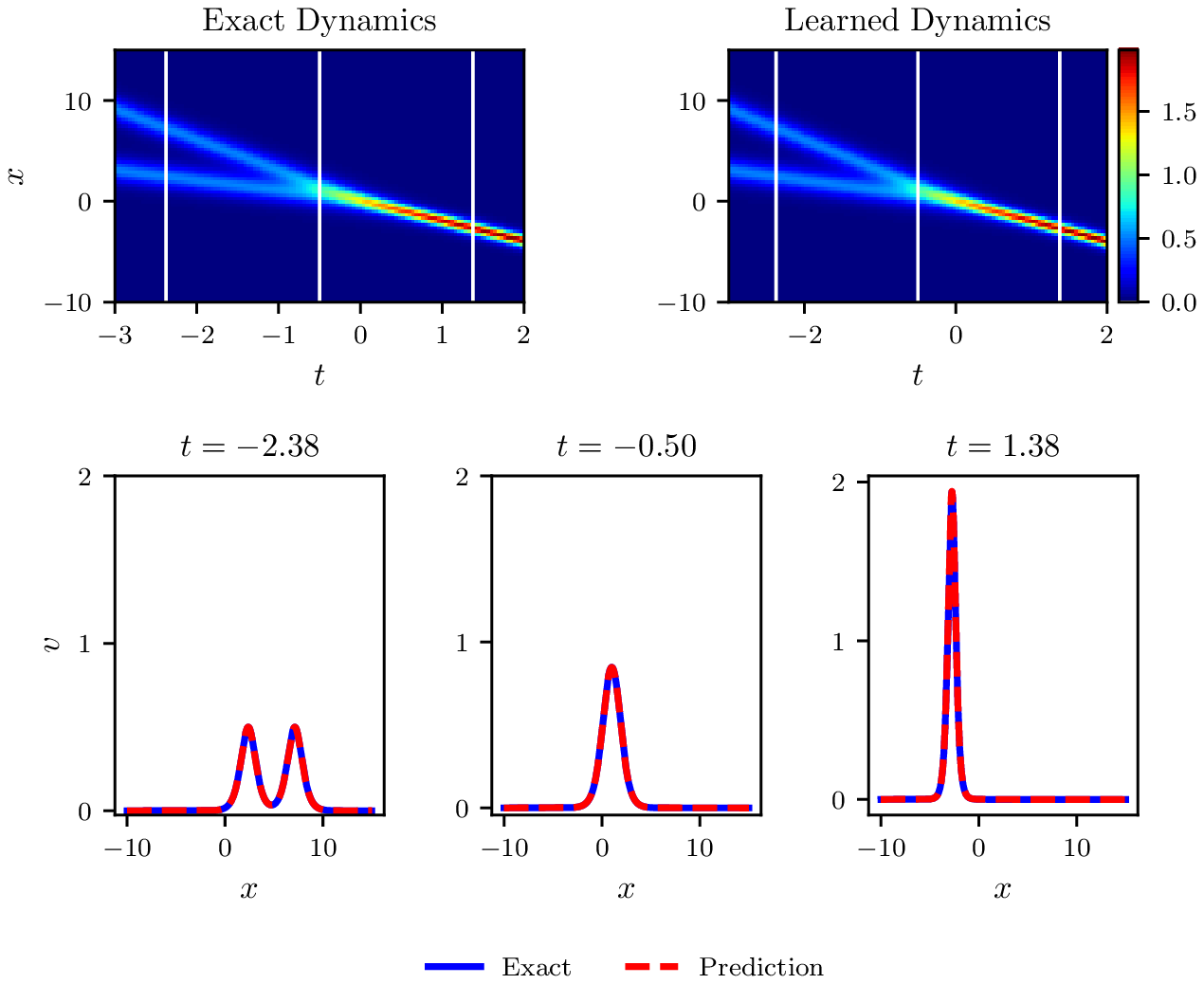}
$c$
\includegraphics[width=7cm,height=5cm]{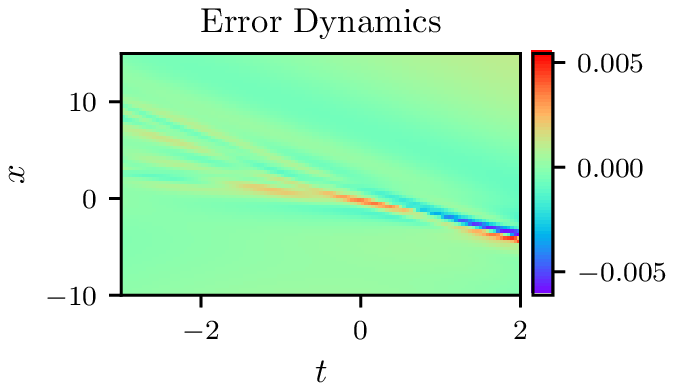}
$d$
\caption{(Color online) Interaction solution $u(x,t)$ and $v(x,t)$ between soliton and resonance of the classical Boussinesq-Burgers equations by two-stage PINN based on conserved quantities: (a) The density diagrams and comparison between the predicted solutions and exact solutions at the three temporal snapshots of $u(x,t)$; (b) The error density diagram of $u(x,t)$; (c) The density diagrams and comparison between the predicted solutions and exact solutions at the three temporal snapshots of $v(x,t)$; (d) The error density diagram of $v(x,t)$.}
\end{figure}

\begin{figure}
\centering
\includegraphics[width=6.5cm,height=5cm]{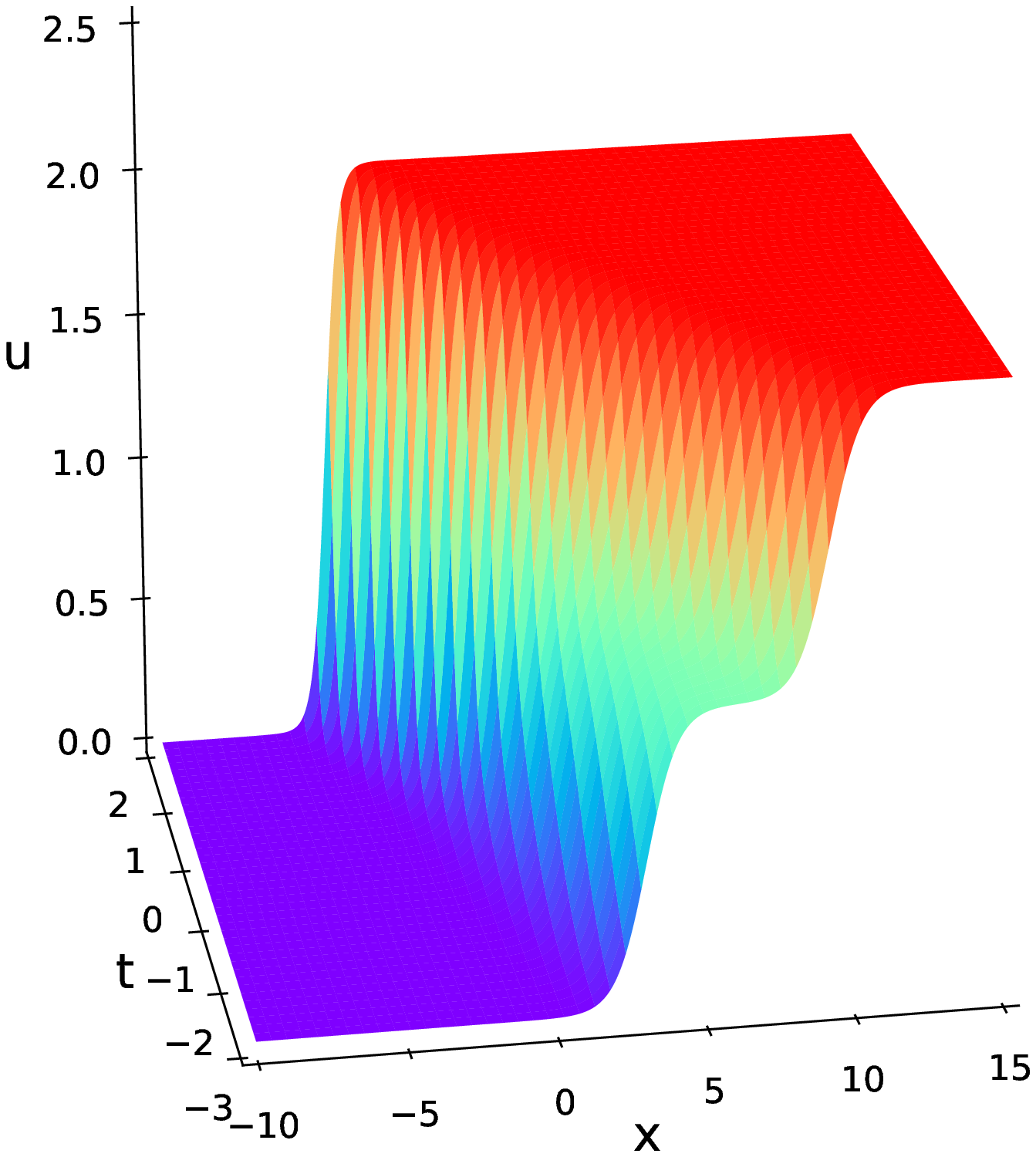}
$a$
\includegraphics[width=6.5cm,height=5cm]{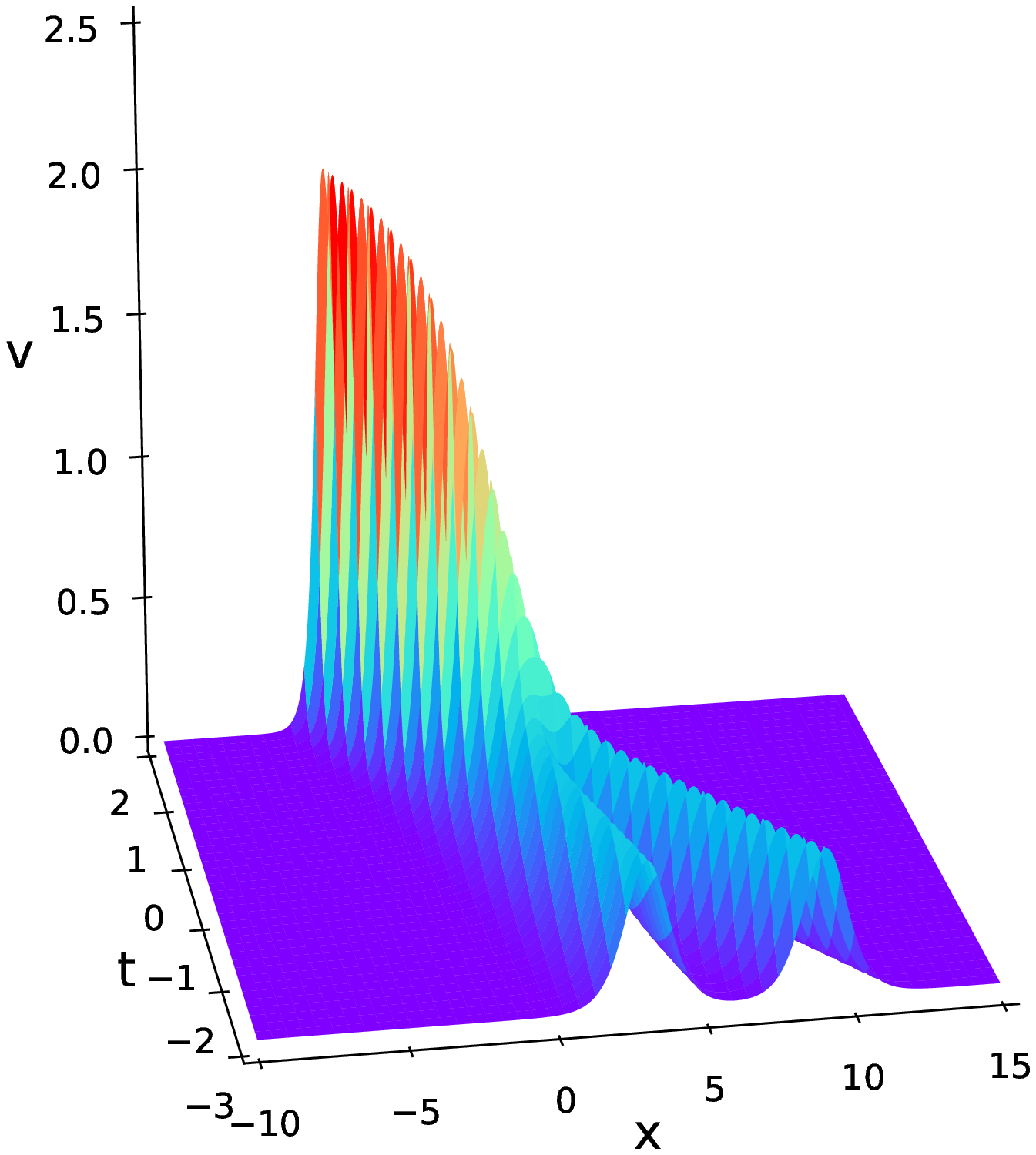}
$b$
\caption{(Color online) Interaction solution $u(x,t)$ and $v(x,t)$ between soliton and resonance of the classical Boussinesq-Burgers equations by two-stage PINN based on conserved quantities: (a) The three-dimensional plot of $u(x,t)$;
(b) The three-dimensional plot of $v(x,t)$.}
\end{figure}

In stage one, we use the original PINN method. After 784 times iterations in about 267.7466 seconds, the relative $\mathbb{L}_2$ error of $u$ is 3.536702e-04 and that of $v$ is 3.304951e-03. In stage two, where the conserved quantity is considered, the relative $\mathbb{L}_2$ error of $u$ is 2.756669e-04 and that of $v$ is 2.576679e-03 after 6411 times iterations in about 2421.5709 seconds. To compare the performance of two methods, the results of calculation shows that the error reduction rate ($ERR$) of $u$ is $22.06\%$ and that of $v$ is $22.04\%$, which are shown in Table. 2. Compared with the original PINN, we also confirm that the precision and the generalization ability of neural networks can be improved by our two-stage PINN method based on conserved quantities.

\begin{table}[htbp]
  \caption{Interaction solution of the classical Boussinesq-Burgers equations: relative $\mathbb{L}_2$ errors of PINN and two-stage PINN based on conserved quantities as well as error reduction rates.}

  \centering
  \begin{tabular}{p{3.38cm}|p{3cm}p{4cm}p{4cm}}
  \toprule
  \textbf{\diagbox[height=0.7cm]{Solution}{Method}} &\textbf{\quad PINN} & \textbf{Two-stage PINN} & \textbf{Error reduction rate} \\
  \midrule
  \textbf{\quad\quad\quad\quad u}   &3.536702e-04&\quad\quad2.756669e-04&\quad\quad\quad22.06\%\\
  \textbf{\quad\quad\quad\quad v}   &3.304951e-03&\quad\quad2.576679e-03&\quad\quad\quad22.04\%\\
  \bottomrule
  \end{tabular}
\end{table}

\section{Data-driven soliton molecule and new types of solitons of the Sawada-Kotera equation}
In recent years, soliton molecules, bound states of solitons, have been widely concerned. A pair of bright solitons, bound together by a dark soliton were discovered in optical fibers through numerical simulations and experimental verifications in 2005 \cite{Stratmann2005}. Later, soliton molecules were obtained in dipolar Bose-Einstein condensates by the method of numerical prediction \cite{Lakomy2012}. Lou \cite{Lou2020} used a new mechanism, namely the velocity resonant, to find soliton molecules in three fifth order integrable systems (fifth order KdV, KK and SK equations). Ren et al. \cite{Ren2020} studied soliton molecules of the Korteweg-de Vries equation with higher-order corrections via the velocity resonance mechanism and they found the collision between a soliton molecule and one soliton is elastic.

To our knowledge, there is poor study of the data-driven soliton molecules by physics-informed neural networks. Consequently, in this part, abundant travelling wave structures are numerically simulated through the two-stage PINN method based on conserved quantities, like the soliton molecule, kink-antikink molecule and so on.

For the following Sawada-Kotera (SK, also called Caudrey-Dodd-Gibbon-Sawada-Kotera (CDGSK)) equation
\begin{align}\label{eq4.1}
u_{t}+u_{x x x x x}+15 u u_{x x x}+45 u^{2} u_{x}+15 u_{x} u_{x x}=0,	
\end{align}
which was introduced in Ref. \cite{Sawada1974, Caudrey1976}, Ye et al. \cite{Ye2004} obtained many new periodic travelling wave solutions via Jacobi elliptic function linear superposition approach. Meanwhile, singular travelling wave solutions of this equation were researched \cite{Bilge1996}. Lou \cite{Lou1993} derived the inverse recursion operator by using the pseudopotential of SK.

In this part, we aim to reproduce dynamic behaviors of the soliton molecule and new types of solitons for Eq. \eqref{eq4.1}. Wang et al.  \cite{WangYao2020} obtained the soliton molecule solutions via the travelling wave approach
\begin{align}\label{eq4.2}
u=-a k^{2}+6 a c \frac{c+\cosh \left[\sqrt{3 a} k\left(x-9 a^{2} k^{4} t-b\right)\right]}{\left[c \cosh \left[\sqrt{3 a} k\left(x-9 a^{2} k^{4} t-b\right)\right]+1\right]^{2}},
\end{align}
where $a>0$, $c, k$ and $b$ are arbitrary constants. Here, we consider the Sawada-Kotera (SK) equation with the first kind of boundary condition (Dirichlet boundary condition) as follows
\begin{equation}\label{eq4.3}
\begin{split}
\begin{cases}
u_{t}+u_{x x x x x}+15 u u_{x x x}+45 u^{2} u_{x}+15 u_{x} u_{x x}=0,x\in[x_0,x_1],t\in[t_0,t_1],\\
u(x,t_0)=u_0(x),\\
u(x_0,t)=a_1(t),\\
u(x_1,t)=a_2(t).
\end{cases}
\end{split}
\end{equation}
Obviously, the governing equation $f(x,t)$ is given by
\begin{align}\label{eq4.4}
f:=	u_{t}+u_{x x x x x}+15 u u_{x x x}+45 u^{2} u_{x}+15 u_{x} u_{x x},
\end{align}
and $u$ is a conserved density considered in this section.

We mainly show the following soliton molecule and new types of solitons of the Sawada-Kotera equation:

\subsection{The soliton molecule (SM) for $0 <c \ll \frac{1}{2}$}
\quad

After taking $c=\frac{1}{4000}, k=a=1, b=0$, $[x_0,x_1]=[-15,15], [t_0,t_1]=[-0.01,0.01]$, we have:
\begin{align}
&u(-15,t)=-1+\frac{3\left(\frac{1}{4000}+\cosh (\sqrt{3}(-9 t-15))\right)}{2000\left(\frac{\cosh (\sqrt{3}(-9 t-15))}{4000}+1\right)^{2}},\nonumber\\
&u(15,t)=-1+\frac{3\left(\frac{1}{4000}+\cosh (\sqrt{3}(-9 t+15))\right)}{2000\left(\frac{\cosh (\sqrt{3}(-9 t+15))}{4000}+1\right)^{2}},\nonumber\\
&u_0(x)=-1+3\frac{\frac{1}{4000}+\cosh(\sqrt{3}(0.09+x))}{2000 (\frac{1}{4000}\cosh(\sqrt{3}(0.09+x))+1)^2}.
\end{align}

With the aid of MATLAB, we divide spatial region $[x_0,x_1]=[-15,15]$ into $N_x=513$ discrete equidistance points and time region $[t_0,t_1]=[-0.01,0.01]$ into $N_t=201$ discrete equidistance points. Thus, the solution $u$ in the given spatiotemporal domain is discretized into $513 \times 201$ data points. We randomly select $N_u=100$ points $\{x^i_u,t^i_u,u^i\}^{N_u}_{i=1}$ from the initial-boundary dataset and proceed by sampling $N_f=2000$ collocation points $\{x_{f}^i,t_{f}^i\}^{N_{f}}_{i=1}$ via the Latin hypercube sampling method. A 9-layer feedforward neural network with 40 neurons per hidden layer is constructed to learn the soliton molecule (SM) of the Sawada-Kotera equation. Besides, we use the hyperbolic tangent ($tanh$) activation function and initialize weights of the neural network with the Xavier initialization. The derivatives of the network $u$ with respect to time $t$ and space $x$ are derived by automatic differentiation.

The loss function of stage one is \eqref{E3} and that of stage two is \eqref{E6} where we choose $N_s=2000, N_c=20$ and $MSE_m$ is given by
\begin{align}
MSE_m&=\frac{1}{N_c}\sum^{N_c}_{i=1}|m(t_m^i)-m(t_0)|^2	\nonumber\\
&\approx \frac{1}{N_c}\sum^{N_c}_{i=1}\bigg|\sum^{N_x}_{j=2} \Widehat{u}(x^j, t_m^i) \frac{x_1-x_0}{N_x-1}-\sum^{N_x}_{j=2} u(x^j, t_0) \frac{x_1-x_0}{N_x-1}\bigg|^2.
\end{align}
where $u(x^j, t_0)$ and $\Widehat{u}(x^j, t_m^i)$ represent the true value and predictive value, respectively.
Then the L-BFGS algorithm is utilized to optimize loss functions above.

The numerical solution can be obtained through our two-stage PINN method. When the original PINN method is applied in stage one, it achieves the relative $\mathbb{L}_2$ error of 4.841810e-03 after 10736 times iterations in about 5551.9234 seconds. In stage two, the conserved quantity ($m=\int_{x_0}^{x_1} u {\rm dx} \approx \sum^{N_x}_{j=2} u(x^j, t) \frac{x_1-x_0}{N_x-1}$) is considered. After 7107 times iterations in about 3987.8433 seconds, the relative $\mathbb{L}_2$ error of $u$ is 4.336727e-03.

Fig. 7 exhibits the density diagrams of soliton molecule $u(x,t)$ and comparison between the predicted solutions and exact solutions at different time points $t=-0.01, 0, 0.01$. It is obvious that dynamic behavior of this solution can be well simulated with high precision from contrast in the (a) of Fig. 7.

\begin{figure}
\centering
\includegraphics[width=8cm,height=5cm]{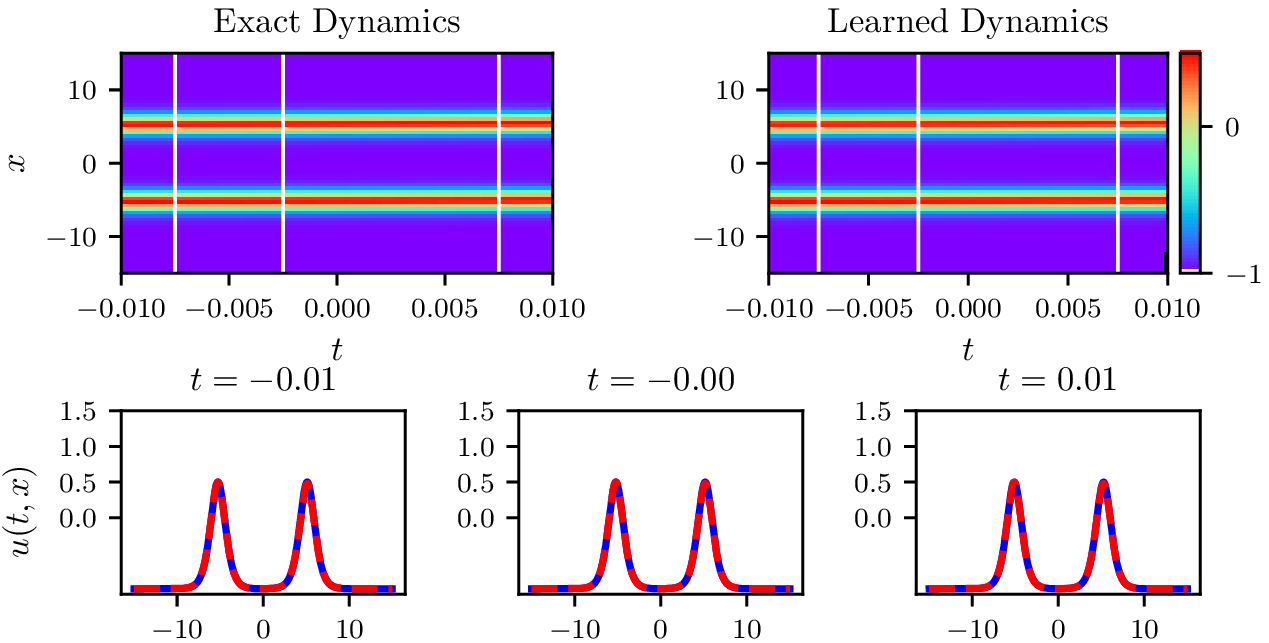}
$a$
\includegraphics[width=8cm,height=5cm]{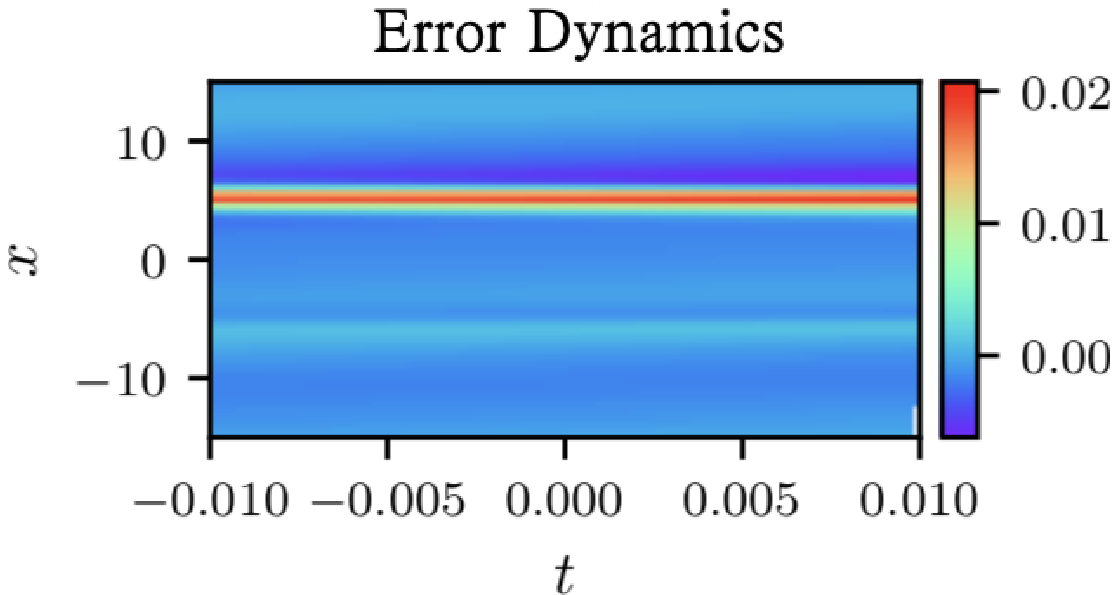}
$b$
\caption{(Color online) Soliton molecule $u(x,t)$ of the Sawada-Kotera equation by two-stage PINN based on conserved quantities: (a) The density diagrams and comparison between the predicted solutions and exact solutions at the three temporal snapshots of $u(x,t)$;
(b) The error density diagram of $u(x,t)$.}
\end{figure}

\subsection{The M-shape double-peak soliton for $c< \frac{1}{2}$}
\quad

Here we take $c=\frac{1}{4}, k=a=1, b=0$, $[x_0,x_1]=[-8,8], [t_0,t_1]=[-0.01,0.01]$, then the initial-boundary conditions are obtained
\begin{align}
&u(-8,t)=-1+\frac{3\left(\frac{1}{4}+\cosh (\sqrt{3}(-9 t-8))\right)}{2\left(\frac{\cosh (\sqrt{3}(-9 t-8))}{4}+1\right)^{2}},\nonumber\\
&u(8,t)=-1+\frac{3\left(\frac{1}{4}+\cosh (\sqrt{3}(-9 t+8))\right)}{2\left(\frac{\cosh (\sqrt{3}(-9 t+8))}{4}+1\right)^{2}},\nonumber\\
&u_0(x)=-1+3\frac{\frac{1}{4}+\cosh(\sqrt{3}(0.09+x))}{2 (\frac{1}{4}\cosh(\sqrt{3}(0.09+x))+1)^2}.
\end{align}

Then, we construct a 7-layer feedforward neural network with 40 neurons per hidden layer to simulate the M-shape double-peak soliton of the Sawada-Kotera equation. The initial-boundary data is obtained via the same generation and sampling method in Section 4.1 and here we choose $N_x=513,N_t=201$ as well. Moreover, the selected values of $N_u, N_f, N_s,N_c$ and the neural network setting, such as loss functions, the optimization algorithm, the activation function and so on, are the same as the previous section.

By means of the two-stage PINN method based on conserved quantities, we finally acquire the data-driven M-shape double-peak soliton solution.

In stage one, the PINN model achieves the relative $\mathbb{L}_2$ error of 1.557140e-03 after 5294 times iterations in about 1927.8552 seconds. In stage two, where we introduce the conserved quantity into the neural network, after 4565 times iterations in about 1796.5973 seconds, the relative $\mathbb{L}_2$ error of $u$ is 8.148663e-04.

In Fig. 8, we present the density diagrams of M-shape double-peak soliton $u(x,t)$ and comparison between the predicted solutions and exact solutions. Besides, in the (b) and (c) of Fig. 8, error density diagrams generated by the original PINN and two-stage PINN based on conserved quantities are plotted separately, which fully verified the advantage of high precision of our improvement.

\begin{figure}
\centering
\includegraphics[width=5.9cm,height=4cm]{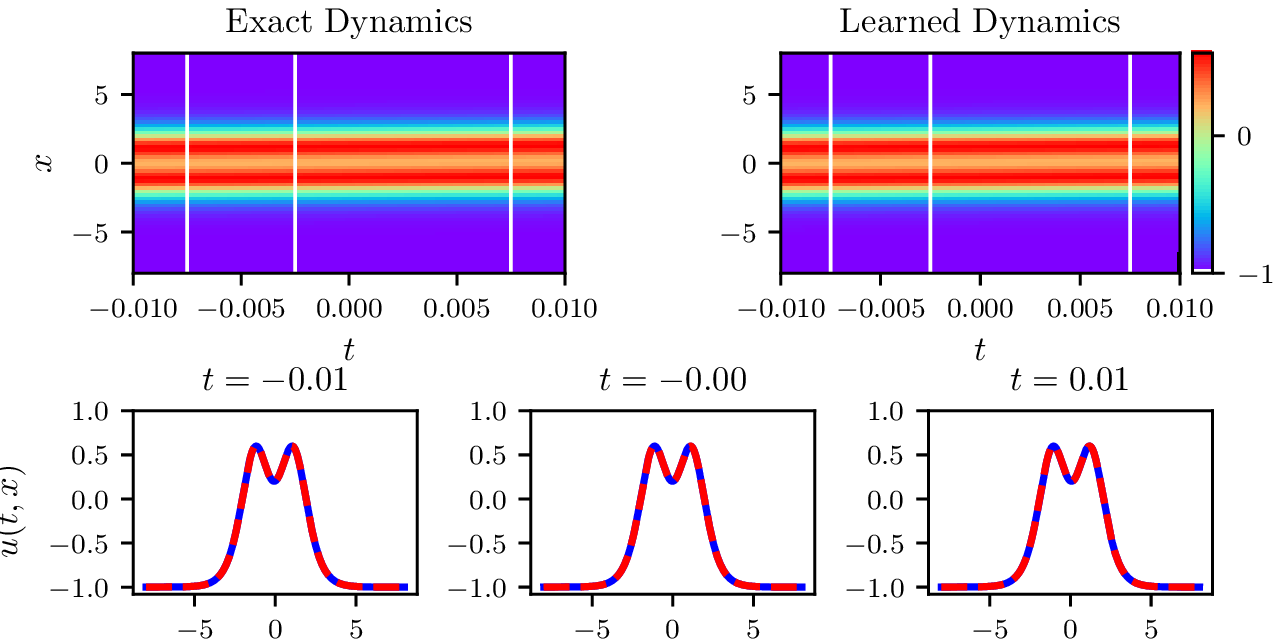}
$a$
\includegraphics[width=5.5cm,height=4cm]{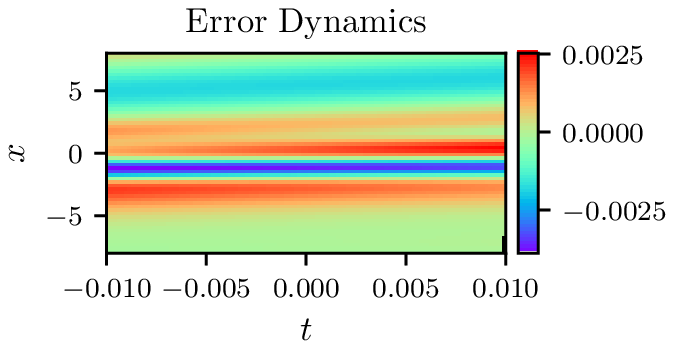}
$b$
\includegraphics[width=5.5cm,height=4cm]{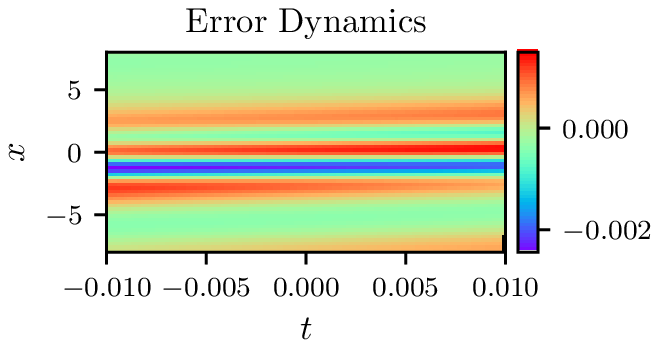}
$c$
\caption{(Color online) M-shape double-peak soliton $u(x,t)$ of the Sawada-Kotera equation: (a) The density diagrams and comparison between the predicted solutions and exact solutions at the three temporal snapshots of $u(x,t)$ by two-stage PINN based on conserved quantities;
(b) The error density diagram of $u(x,t)$ by original PINN;
(c) The error density diagram of $u(x,t)$ by two-stage PINN based on conserved quantities.}
\end{figure}

\subsection{The kink-antikink molecule (KAKM) or plateau soliton for $c=\frac{1}{2}$}
\quad

The initial-boundary conditions are given by
\begin{align}
&u(-6,t)=-1+\frac{3\left(\frac{1}{2}+\cosh (\sqrt{3}(-9 t-6))\right)}{\left(\frac{\cosh (\sqrt{3}(-9 t-6))}{2}+1\right)^{2}},\nonumber\\
&u(6,t)=-1+\frac{3\left(\frac{1}{2}+\cosh (\sqrt{3}(-9 t+6))\right)}{\left(\frac{\cosh (\sqrt{3}(-9 t+6))}{2}+1\right)^{2}},\nonumber\\
&u_0(x)=-1+3\frac{\frac{1}{2}+\cosh(\sqrt{3}(0.09+x))}{(\frac{\cosh(\sqrt{3}(0.09+x))}{2}+1)^2},
\end{align}
after choosing corresponding parameters as $c=\frac{1}{2}, k=a=1, b=0$, $[x_0,x_1]=[-6,6], [t_0,t_1]=[-0.01,0.01]$.

Similarly,  we divide spatial region $[x_0,x_1]=[-6,6]$ into $N_x=513$ discrete equidistance points and time region $[t_0,t_1]=[-0.01,0.01]$ into $N_t=201$ discrete equidistance points. Thus, the solution $u$ is discretized into $513 \times 201$ data points in the given spatiotemporal domain. The initial-boundary dataset sampled randomly is served as input to the neural network for training. In the process of establishing the two-stage physics-informed neural network, we also adopt the fully-connected structure with the Xavier initialization and hyperbolic tangent activation function. The loss functions of two stages are optimized in the same way as described above. The only difference is the number of hidden layers. We construct a 8-layer feedforward neural network with 40 neurons per hidden layer here.

After 5381 times iterations in about 2310.8078 seconds, the relative $\mathbb{L}_2$ error of $u$ is 1.766738e-03 in stage one. The numerical results show that after 2972 times iterations, our two-stage PINN model based on conserved quantities achieves the relative $\mathbb{L}_2$ error of 3.678135e-04 in about 1315.2801 seconds in stage two.

Fig. 9 exhibits the density diagrams of the plateau soliton $u(x,t)$ and comparison between the predicted solutions and exact solutions at different time points $t=-0.01, 0, 0.01$. Similarly, error density diagrams generated by the original PINN and two-stage PINN based on conserved quantities are plotted respectively in the (b) and (c) of Fig. 9. Most notably, the relative $\mathbb{L}_2$ of $u(x,t)$ by two-stage PINN based on conserved quantities nearly reach to 5e-4, about one order of magnitude lower than that by the original PINN. This result accentuates that the improved method can enhance the performance in terms of accuracy.

\begin{figure}
\centering
\includegraphics[width=5.9cm,height=4cm]{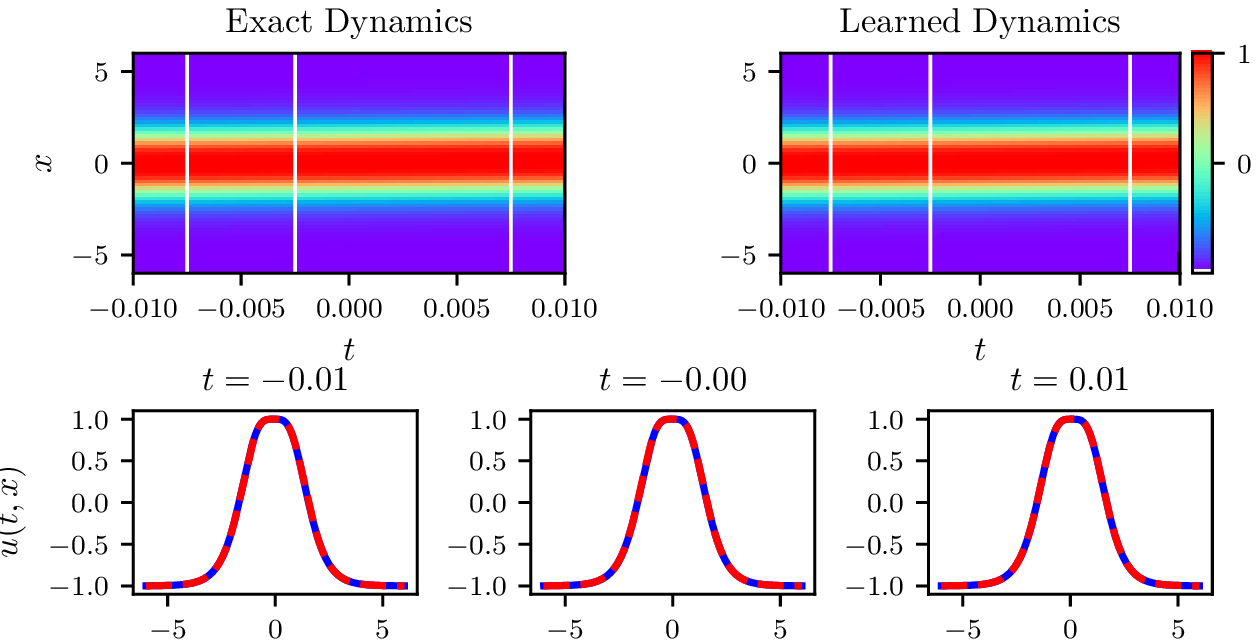}
$a$
\includegraphics[width=5.5cm,height=4cm]{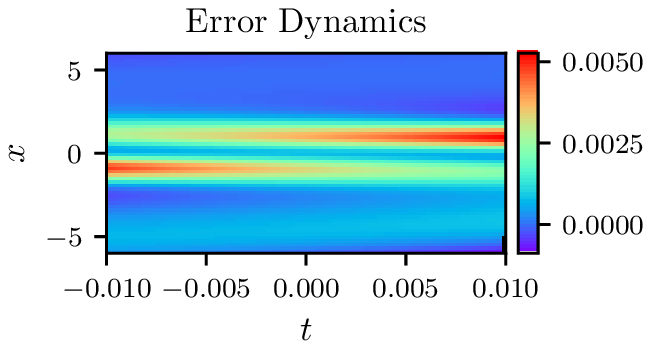}
$b$
\includegraphics[width=5.5cm,height=4cm]{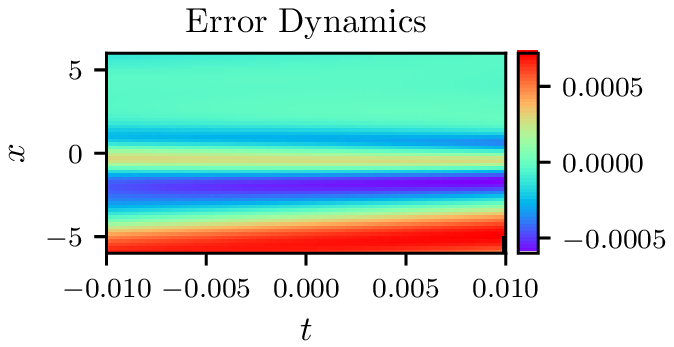}
$c$
\caption{(Color online) Plateau soliton $u(x,t)$ of the Sawada-Kotera equation: (a) The density diagrams and comparison between the predicted solutions and exact solutions at the three temporal snapshots of $u(x,t)$ by two-stage PINN based on conserved quantities;
(b) The error density diagram of $u(x,t)$ by original PINN;
(c) The error density diagram of $u(x,t)$ by two-stage PINN based on conserved quantities.}
\end{figure}

\subsection{The single-peak soliton for $c>\frac{1}{2}$}

We choose $c=1, k=a=1, b=0$, $[x_0,x_1]=[-6,6], [t_0,t_1]=[-0.01,0.01]$, which yields
\begin{align}
&u(-6,t)=-1+\frac{6+6 \cosh (\sqrt{3}(-9 t-6))}{(\cosh (\sqrt{3}(-9 t-6))+1)^{2}},\nonumber\\
&u(6,t)=-1+\frac{6+6 \cosh (\sqrt{3}(-9 t+6))}{(\cosh (\sqrt{3}(-9 t+6))+1)^{2}},\nonumber\\
&u_0(x)=-1+6\frac{1+\cosh(\sqrt{3}(0.09+x))}{(\cosh(\sqrt{3}(0.09+x))+1)^2}.
\end{align}

After exploiting the same data discretization and sampling method, spatial region $[x_0,x_1]=[-6,6]$ and time region $[t_0,t_1]=[-0.01,0.01]$ are divided into $N_x=513$ and  $N_t=201$ discrete equidistance points, respectively. Based on the training data sub-sampled by the the Latin hypercube sampling method, a 8-layer feedforward neural network with 40 neurons per hidden layer is established to derive numerical solutions in the form of single-peak soliton after setting up parameters as $N_u=100, N_f=2000$. Likewise, we employ the Xavier initialization and hyperbolic tangent ($tanh$) activation function as well as the L-BFGS algorithm to optimize loss functions where $N_s=2000, N_c=20$.

With the advantage of the proposed two-stage PINN method, we finally reproduce the single-peak soliton solution.

In stage one, we construct the original PINN model. After 1583 times iterations in about 813.6134 seconds, the relative $\mathbb{L}_2$ error of $u$ is 1.772205e-02. In stage two, where the conserved quantity is considered, the relative $\mathbb{L}_2$ error of $u$ is 9.931406e-03 after 1034 times iterations in about 611.5201 seconds.

In Fig. 10, the density diagrams of single-peak soliton $u(x,t)$ and comparison between the predicted solutions and exact solutions are displayed. Through comparing error density diagrams of two methods showed in the (b) and (c) of Fig. 10, it demonstrates that our two-stage PINN method based on conserved quantities is also more accurate for simulating single-peak soliton.

\begin{figure}
\centering
\includegraphics[width=5.9cm,height=4cm]{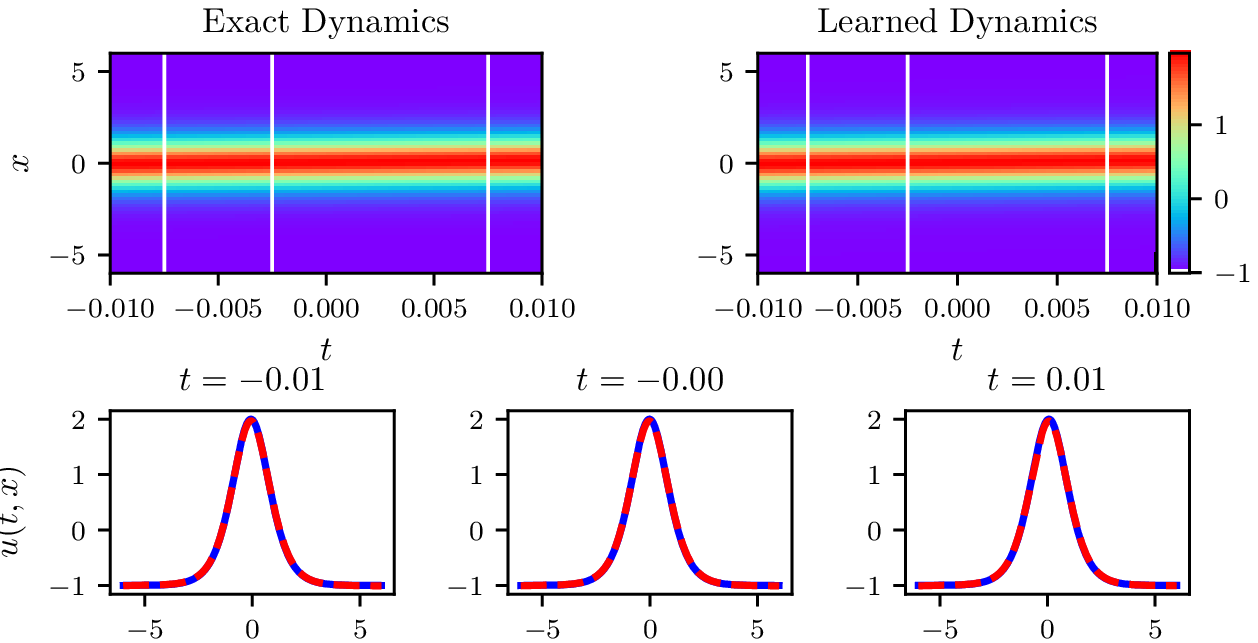}
$a$
\includegraphics[width=5.5cm,height=4cm]{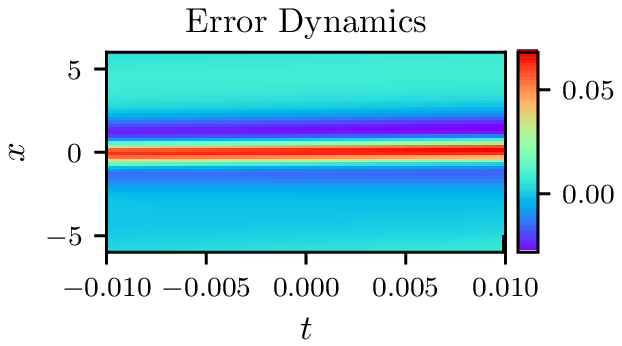}
$b$
\includegraphics[width=5.5cm,height=4cm]{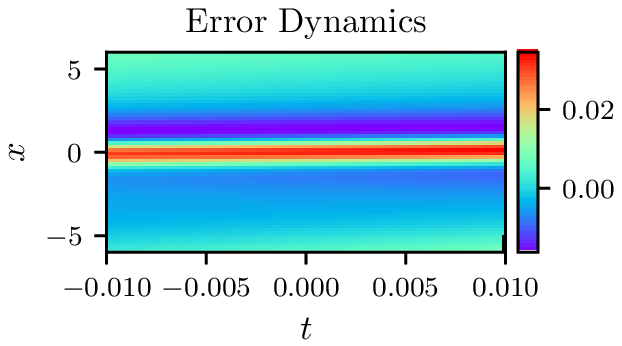}
$c$
\caption{(Color online) Single-peak soliton $u(x,t)$ of the Sawada-Kotera equation: (a) The density diagrams and comparison between the predicted solutions and exact solutions at the three temporal snapshots of $u(x,t)$ by two-stage PINN based on conserved quantities;
(b) The error density diagram of $u(x,t)$ by original PINN;
(c) The error density diagram of $u(x,t)$ by two-stage PINN based on conserved quantities.}
\end{figure}

In Fig. 11, the three-dimensional plots of four structures are plotted respectively: soliton molecule, M-shape double-peak soliton, plateau soliton and single-peak soliton. It illustrates that the two-stage PINN method can effectively reproduce different dynamic behaviors.

\begin{figure}
\centering
\includegraphics[width=6cm,height=4cm]{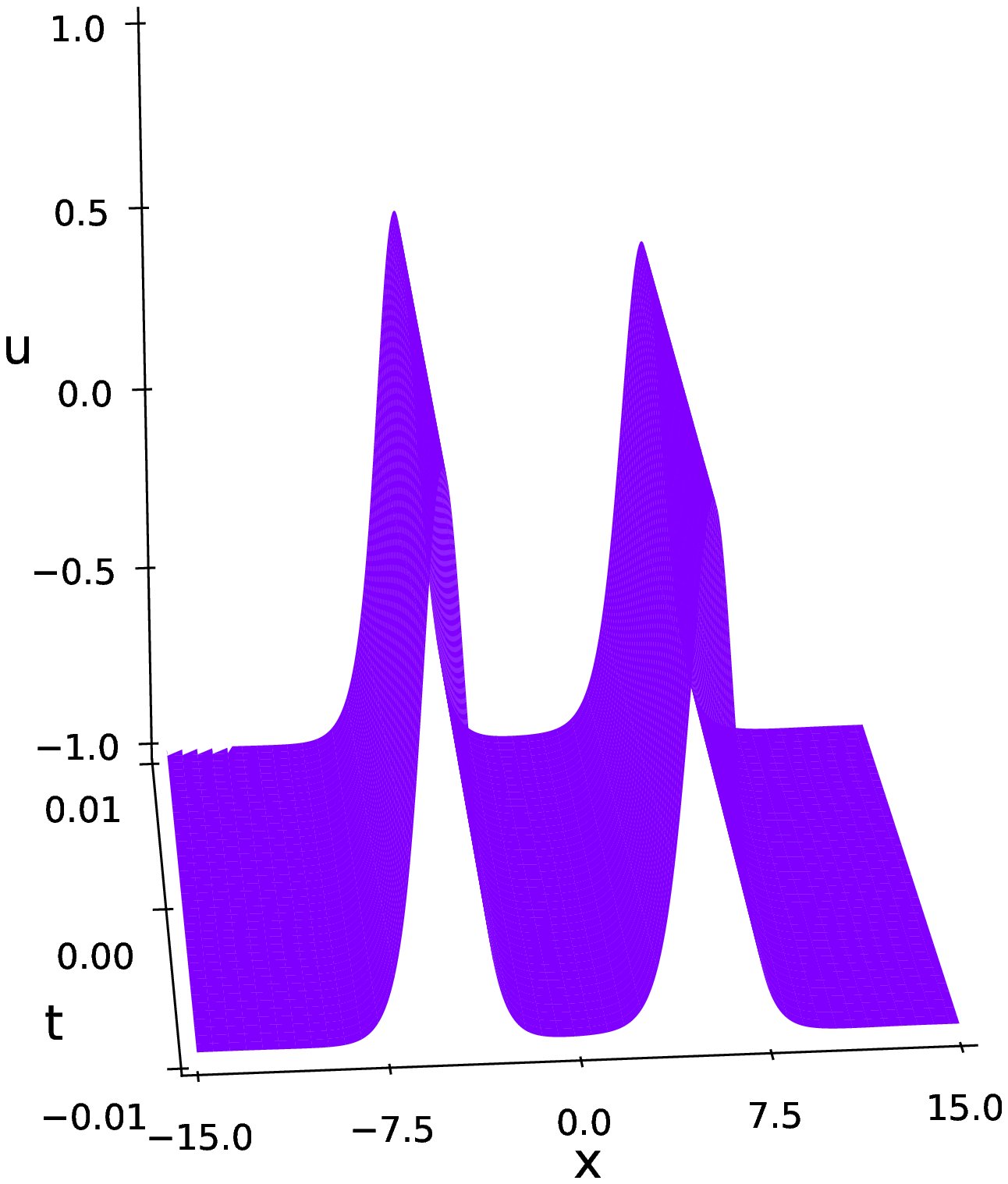}
$a$
\includegraphics[width=6cm,height=4cm]{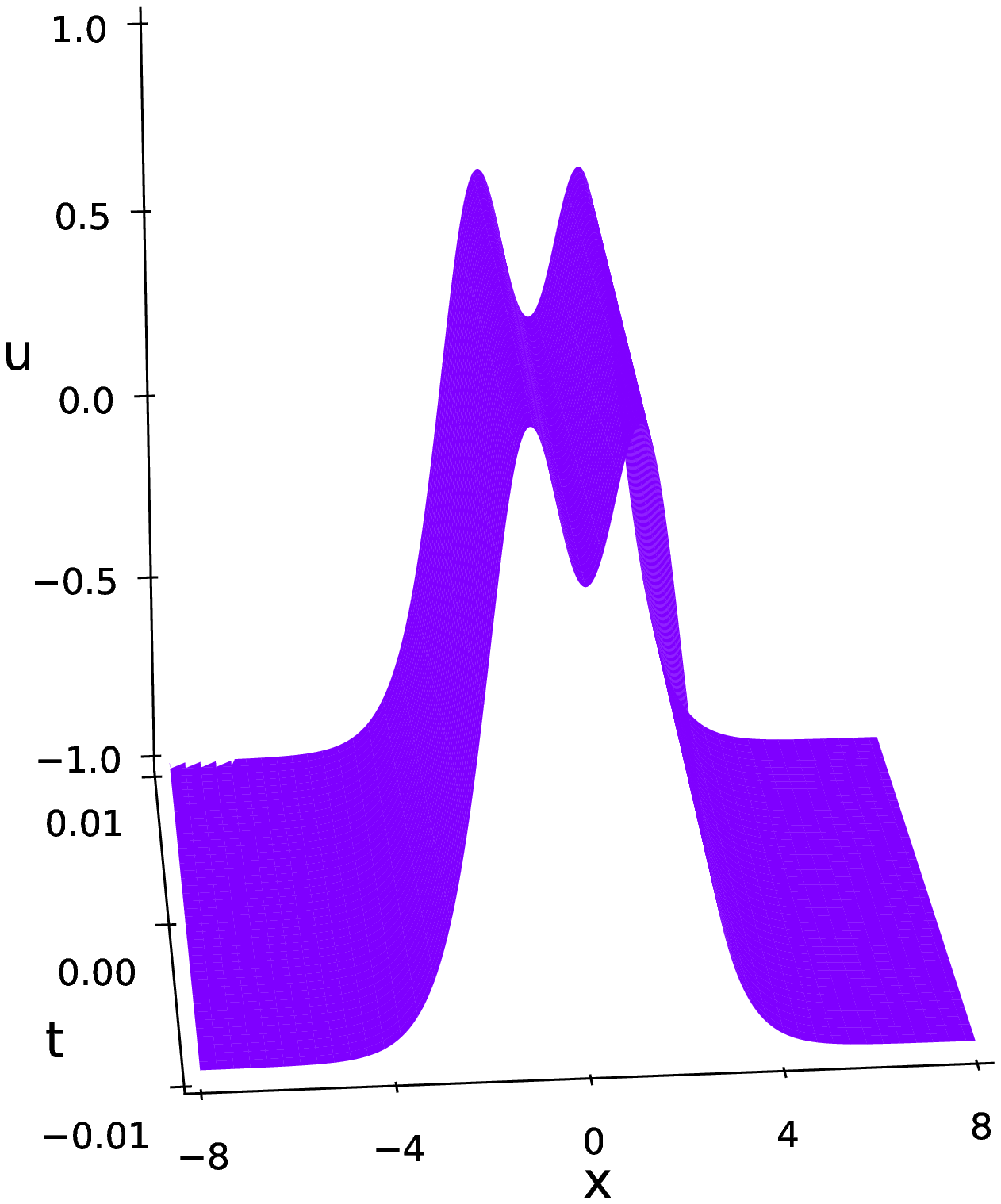}
$b$\\
\includegraphics[width=6cm,height=4cm]{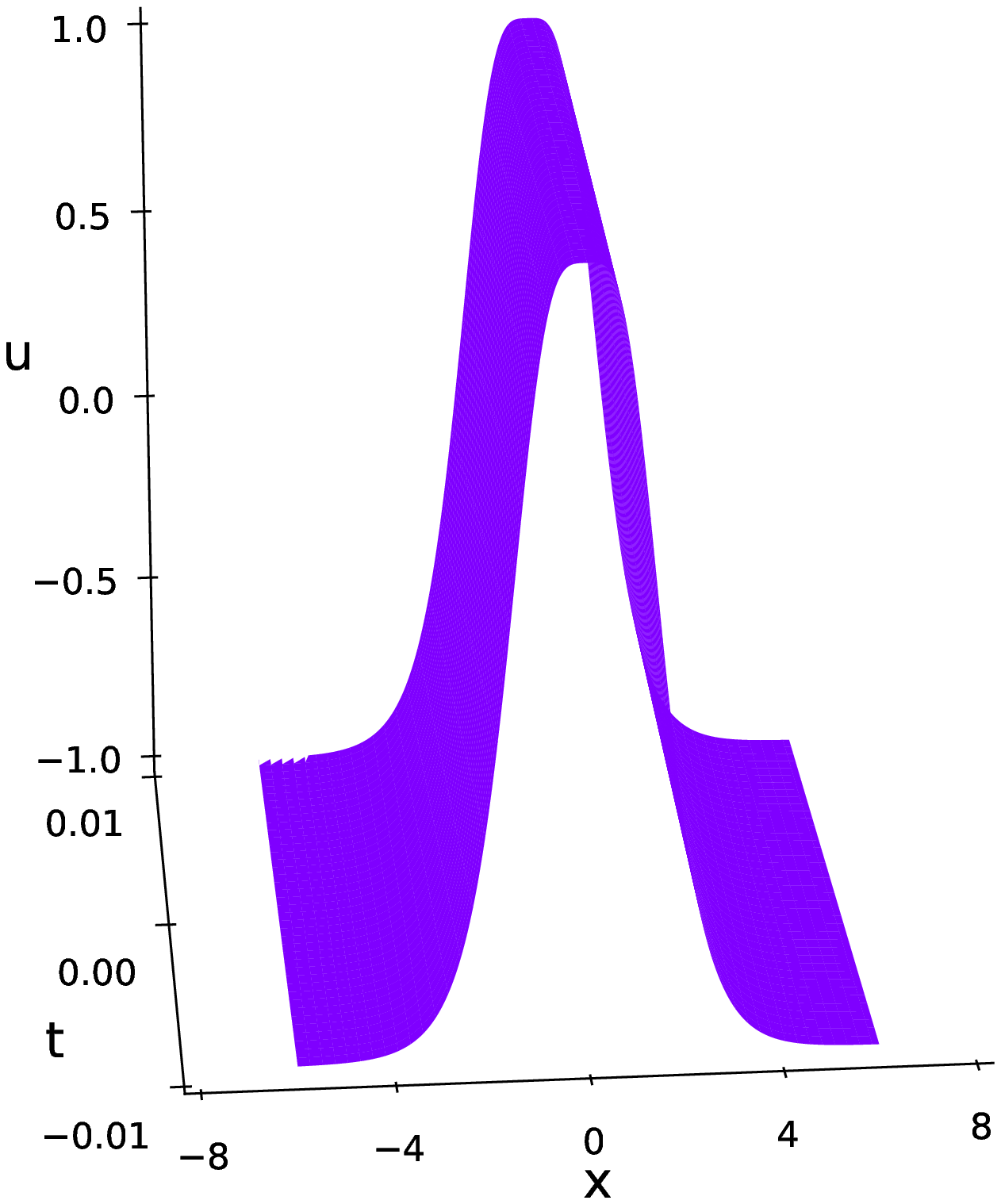}
$c$
\includegraphics[width=6cm,height=4cm]{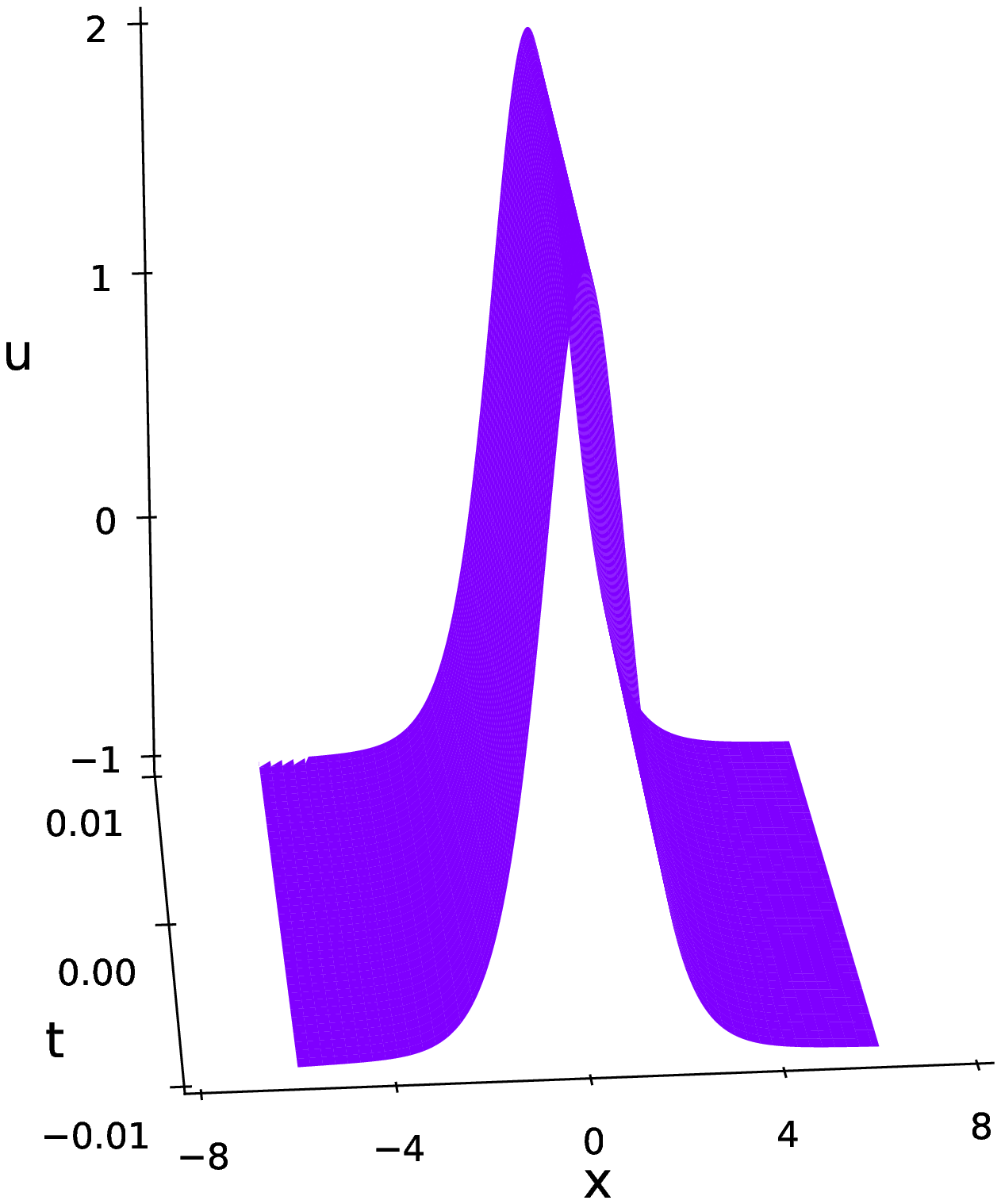}
$d$
\caption{(Color online) The three-dimensional plots of background induced soliton and soliton molecules $u(x,t)$ by two-stage PINN based on conserved quantities: (a) Soliton molecule; (b) M-shape double-peak soliton; (c) Plateau soliton; (d) Single-peak soliton.}
\end{figure}

Similarly, according to the relative $\mathbb{L}_2$ error of the PINN method ($RE_1$) and that of the two-stage PINN method based on conserved quantities ($RE_2$), the error reduction rate ($ERR$) can be obtained. In addition, Table. 3 shows the relative $\mathbb{L}_2$ errors of above solutions as well as the constrast of two methods in terms of error reduction rates.

\begin{table}[htbp]
  \caption{Soliton molecule and new types of solitons of the Sawada-Kotera equation: relative $\mathbb{L}_2$ errors of PINN and two-stage PINN based on conserved quantities as well as error reduction rates.}

  \centering
  \begin{tabular}{p{3.5cm}|p{3cm}p{4cm}p{4cm}}
  \toprule
  \textbf{\diagbox[height=0.7cm]{Solution}{Method}} &\textbf{\quad PINN} &\textbf{Two-stage PINN} &\textbf{Error reduction rate} \\
  \midrule
  \textbf{Soliton molecule}   &4.841810e-03&\quad\quad4.336727e-03&\quad\quad\quad10.43\%\\
  \textbf{M-shape soliton}   &1.557140e-03&\quad\quad8.148663e-04&\quad\quad\quad47.67\%\\
  \textbf{Plateau soliton}   &1.766738e-03&\quad\quad3.678135e-04&\quad\quad\quad79.18\%\\
  \textbf{Single-peak soliton}  &1.772205e-02&\quad\quad9.931406e-03&\quad\quad\quad43.96\%\\
  \bottomrule
  \end{tabular}
\end{table}

From Table. 3, it can be seen that the proposed two-stage PINN method based on conserved quantities remarkably improves the original PINN method according to error reduction rates. Especially for the plateau soliton, its error reduction rate (79.18\%) is extraordinarily significant in the numerical experiments. Consequently, our improvement is shown to effectively enhance the prediction accuracy.

\section{Conclusion}

In this paper, we aim to devise a more targeted PINN algorithm tailored to the nature of equations by introducing conserved quantities of nonlinear systems into neural networks, which implies that the underlying information of the given equations is dug out to improve the precision and reliability. Moreover, the original PINN method considers the local constraints at certain points solely, which evokes the question of whether we can impose constraints from a global perspective. For these purposes, we propose the two-stage PINN method based on conserved quantities. In stage one, the original PINN is applied. In stage two, we additionally introduce the measurement of conserved quantities into mean squared error loss to train neural networks to achieve further optimization of the numerical solution in the first stage. This methodology provides a promising new direction to devise deep learning algorithms with the advantages of integrable systems.

At the same time, we richly exemplify the use of this improved PINN method by simulating the one-soliton solution of the Boussinesq-Burgers equations as well as the interaction solution between a soliton and one resonant of the classical Boussinesq-Burgers equations. Besides, considering that there is poor study of the data-driven soliton molecules by physics-informed neural networks, we reproduce the dynamical behaviors of the soliton molecule, M-shape double-peak soliton, plateau soliton and single-peak soliton for the Sawada-Kotera (SK) equation.

For the sake of comparing the performances of two methods: the original PINN and  two-stage PINN based on the conserved quantities, we calculate error reduction rates according to their own relative $\mathbb{L}_2$ errors. Remarkably, results indicate that two-stage PINN method based on conserved quantities can obviously improve prediction accuracy and enhance the ability of generalization, which implies that our improvement is meaningful in simulating solutions of nonlinear partial differential equations. Meanwhile, this is the first time that features of integrable systems are introduced to PINN method. Thus, our practice can solve partial differential equations much more pertinently and promote the development of this field.

However, our proposed method increases the training cost for improving the accuracy. In the future, we  will devote to devise a new physics-informed neural network algorithm which can improve prediction accuracy and generalization ability without sacrificing efficiency.

\section*{Acknowledgments}
The authors would like to thank Zhengwu Miao sincerely for providing with support and guidance. This work is supported by Global Change Research Program of China (No.2015CB953904), National Natural Science Foundation of China (No.11675054) and Science and Technology Commission of Shanghai Municipality (No.18dz2271000).

\end{document}